# Promise of Commercialization: Carbon Materials for Low-Cost Perovskite Solar Cells*

Yu Cai(蔡宇)[a,b], Lusheng Liang(梁禄生)[b] and Peng Gao(高鹏)[b†]

[1] *Department of Chemistry, Fujian Normal University, 350007, Fuzhou, Fujian P.R.China*

[2] *Laboratory for Advanced Functional Materials, Xiamen Institute of Rare-earth Materials, Chinese Academy of Science, 361021, Xiamen, Fujian, P.R.China*

Perovskite solar cells (PVSCs) have attracted extensive studies due to their high power conversion efficiency (PCE) with low-cost in both raw material and processes. However, there remain obstacles that hinder the way to its commercialization. Among many drawbacks in PVSCs, we note the problems brought by the use of noble metal counter electrodes (CEs) such as gold (Au) and silver (Ag). The costly Au and Ag need high energy-consumption thermal evaporation process which can be made only with expensive evaporation equipment under vacuum. All the factors elevate the threshold of PVSCs' commercialization. Carbon material, on the other hand, is a readily available electrode candidate for the application as CE in the PVSCs. In this review, endeavors on PVSCs with low-cost carbon materials will be comprehensively discussed based on different device structures and carbon composition. We believe that the PVSCs with carbon-based CE hold the promise of commercialization of this new technology.

## 1. Introduction

Along with the gradual exhaustion of fossil energies and the environmental pollution caused by employing them, people increasingly realize the importance of renewables. Solar energy is the most abundant energy in the world. In order to convert solar energy into electric energy which can be directly used by modern society, various kinds of solar cells were designed. Solar cell technologies are grouped into three generations by Martin Green[1], the first generation usually refers to mono-crystalline silicon and poly-crystalline silicon solar cells,[2-4] the second generation is the family of thin film solar cells (such as cadmium sulphide, cadmium telluride, copper indium diselenide and amorphous silicon), and the third generation include polymer solar cell (PSC), dye-sensitized solar cell (DSSC), quantum dot solar cell (QDSC) and so on, which utilize new materials and nanotechnology. So far, the family of silicon-based solar cells dominates the world's photovoltaic market, but they suffered from drawbacks such as high manufacturing cost, high energy-consumption processes, and severe environmental pollution during the productive process. The so-called second generation

* Project supported by

† Corresponding author. E-mail: peng.gao@fjirsm.ac.cn

solar cell namely thin-film solar cells, also have some shortcomings, for instance, low power conversion efficiency (PCE), scarcity of raw materials and operational instability. The third-generation solar cells show great prospect because of their interesting features including low-cost, less energy consumption and abundant raw materials. Therefore, the third-generation solar cells conform to the trend of the commercial market of the photovoltaic industry.

Among the third-generation solar cells, dye-sensitized solar cell (DSSCs)[5] attracted intensive research interest due to the low-cost materials and ease of fabrication. The upsurge of investigations on perovskite solar cells (PVSCs) sprang from the community of DSSCs. Since the pioneering work of Kojima et al. which introduced $CH_3NH_3PbBr_3$ and $CH_3NH_3PbBr_3$ as light absorbers in a liquid-state perovskite-sensitized solar cell in 2009, PVSCs have allured intensive studies both from the academic world and industrial community[6-9]. With thousands of publications and increasing attempts to commercialization, PVSCs gradually become an almost matured technology during the past eight years.[10]

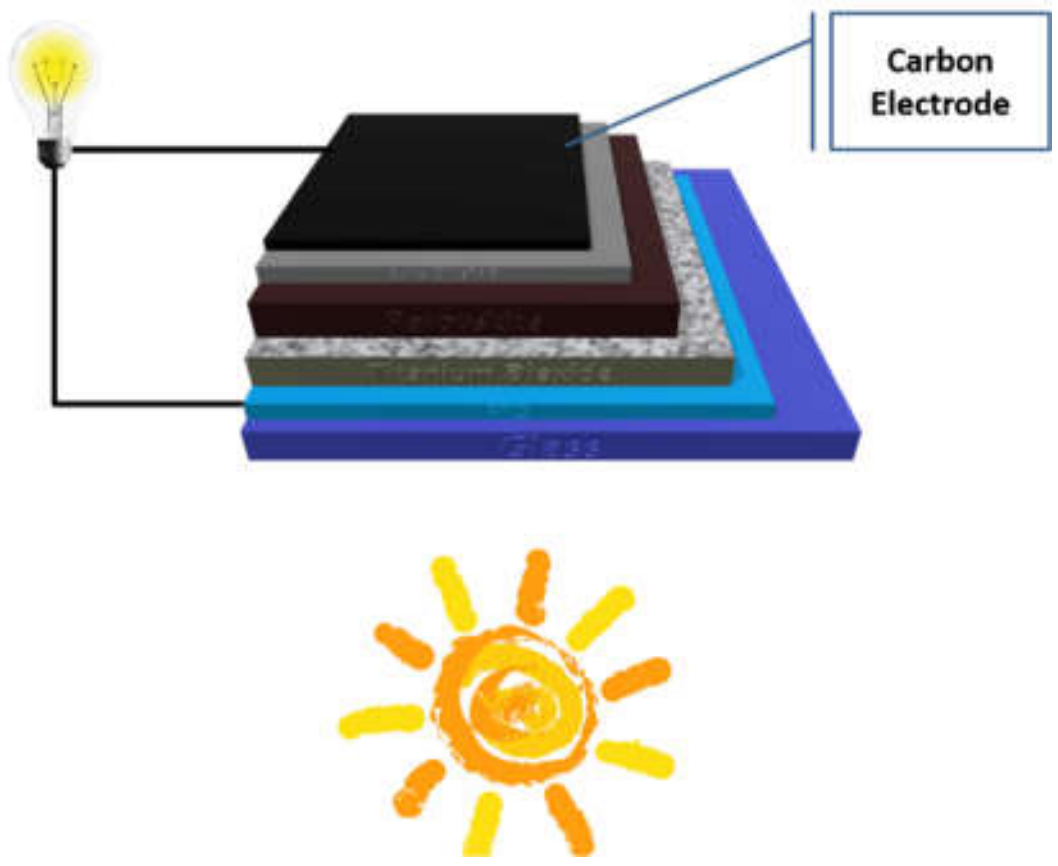

**Fig. 1**. The working Carbon based PVSC under sunlight.

PVSCs usually consist of matrix-supported transparent conductive oxide (TCO) anode, charge selective contact layer (hole/electron transport layer /H(E)TL), perovskite absorbed layer, second charge selective contact layer (electron/hole transport layer) and cathode[11]. PVSCs generate electricity according to the following mechanism (Figure 1): Once illuminated by sunlight, the perovskite layer would absorb photons which have higher energy than its band gap and electron/hole pairs generated in the perovskite layer can be dissociated at the interfaces of perovskite/H(E)TL. Electrons are injected into the conduction band of the ETL and collected by the anode, while holes are transported to the valence band of the HTL and compiled by the cathode. Electrons flow through the external circuit and arrive at the cathode where they reunite with holes.[12] Due to the unique semiconducting and conducting properties of perovskite crystal, such as high optical absorption coefficient, appropriate direct band gap around 1.5eV, ambipolar charge mobility, long carrier lifetime and long carrier diffusion length[13], PVSCs have nowadays obtained a certified efficiency record of 22.1%[14], which is even higher than that of the commercialized polycrystalline silicon solar cells. However, there are still some issues blocking the way towards the commercialization of PVSCs, for instance, instability of the perovskite contacts with moisture and oxygen[15], the high cost of typical hole-transporting-material (HTM) (such as spiro-OMeTAD, P3HT and PTAA) and noble metallic counter electrode (CE) like gold and silver, long-time stability of the cell under ambient condition, some high energy-consumption processes of device fabrication and harsh requirement for fabrication facilities.

Using Au or Ag as the CE is indispensable to fabricate state-of-the-art PVSCs, but Au is too expensive for mass industrial manufacture and evaporating Au or Ag onto the top of the cell is a high energy-consumption process which needs high vacuum and high temperature. Domanski et al. found that, when PVSCs are exposed to a temperature of 70 °C, considerable amounts of gold from the back electrode will diffuse across the HTL into the perovskite material, resulting in dramatic loss of the device performance under working conditions[16]. Ag is cheaper and works well in PVSCs with HTM, but it can become corroded when in contact with the perovskite film, which is likely to be due to the formation of silver halide[17]. Hence, replacing noble metal cathode with low-cost back electrode materials is imperative. Through the unremitting endeavor of many research groups, carbon materials as the cost-efficient choice of back electrodes have been introduced to PVSCs to address issues above. In this review, the application of carbon materials acting as CEs in PVSCs will be exhibited. We will discuss constructions, properties and morphologies of those materials employed as electrodes, and how these characters affect the performance of the corresponding device. A summary of almost all research works using carbon as counter electrode is shown in Table 1.

Table 1. A summary of almost all research works using carbon as counter electrode

| **Electrode material** | **FF** | **$V_{OC}$ (V)** | **$J_{SC}$ (mA·cm$^{-2}$)** | **PCE (%)** | **Device structure** | **Method** | **Ref.** |
|---|---|---|---|---|---|---|---|
| Carbon black/flaky graphite (Carbon paste) | 0.46 | 0.825 | 10.6 | 4.08 | FTO/$TiO_2$ dense layer/mesoporous $TiO_2$/$CH_3NH_3PbI_3$/$ZrO_2$ space layer/Carbon black/graphite | screen printing | [18] |
| Carbon black/spheroidal graphite (Carbon paste) | 0.61 | 0.878 | 12.4 | 6.64 | FTO/$TiO_2$ dense layer/mesoporous$TiO_2$/$CH_3NH_3PbI_3$/$ZrO_2$ space layer/Carbon black/graphite | screen printing | [18] |
| Carbon black | 0.71 | 0.95 | 17.20 | 11.60 | FTO/$TiO_2$ compact layer/$CH_3NH_3PbI_3$/nano Carbon | inkjet printing | [19] |
| Carbon black/graphite (Carbon paste) | 0.57 | 0.953 | 18.73 | 10.20 | FTO/$TiO_2$ dense layer/$TiO_2$ mesoscopic layer/$CH_3NH_3PbI_3$/ mesoscopic Carbon layer/Industrial flexible graphite sheet | screen printing | [20] |
| commercial Carbon paste | 0.55 | 0.90 | 16.78 | 8.31 | FTO/$TiO_2$ compact layer/$TiO_2$ mesoscopic layer/$CH_3NH_3PbI_3$/ commercial Carbon paste | doctor-blading | [21] |
| commercial conductive Carbon paste | 0.54 | 0.80 | 21.02 | 9.08 | FTO/$TiO_2$ compact layer/$TiO_2$ mesoscopic layer/$CH_3NH_3PbI_3$/ Carbon | doctor-blade coating | [22] |
| Carbon black/graphite (Carbon paste) | 0.66 | 0.858 | 22.8 | 12.84 | FTO/$TiO_2$ dense layer/mesoporous $TiO_2$/$CH_3NH_3PbI_3$/$ZrO_2$ layer/Carbon black/graphite | printing | [23] |
| Carbon black/graphite (Carbon paste) | 0.61 | 0.868 | 20.1 | 10.64 | FTO/$TiO_2$ dense layer/mesoporous $TiO_2$/$CH_3NH_3PbI_3$/$ZrO_2$ layer/Carbon black/graphite | screen printing | [24] |
| Carbon black/graphite (Carbon paste) | 0.76 | 0.915 | 21.62 | 15.03 | FTO/$TiO_2$ compact layer/$TiO_2$/$Al_2O_3$/NiO/Carbon (a quadruple- layer)/$CH_3NH_3PbI_3$ | screen-printing | [25] |
| Carbon black/graphite (Carbon paste) | 0.72 | 0.90 | 20.45 | 13.14 | FTO/$TiO_2$ dense layer/mesoporous $TiO_2$/$CH_3NH_3PbI_3$/$ZrO_2$ layer/Carbon black/graphite | screen-printing | [26] |

| Electrode material | FF | $V_{OC}$ (V) | $J_{SC}$ (mA·cm$^{-2}$) | PCE (%) | Device structure | Method | Ref. |
|---|---|---|---|---|---|---|---|
| graphite flake/Carbon black powder (Carbon paste) | 0.63 | 1.002 | 21.30 | 13.53 | FTO/$TiO_2$ dense layer/mesoporous $TiO_2$/$CH_3NH_3PbI_3$/ thermoplastic Carbon film/Al foil | doctor-blading | [27] |
| conductive-Carbon paste | 0.54 | 0.810 | 19.98 | 8.73 | FTO/C-ZnO/$CH_3NH_3PbI_3$/Carbon | doctor-blading | [28] |
| conductive- Carbon paste | 0.42 | 0.76 | 13.38 | 4.29 | ITO/PEN/ZnO/$CH_3NH_3PbI_3$/Carbon | doctor-blading | [28] |
| Carbon black/graphite (Carbon paste) | 0.71 | 0.89 | 18.2 | 11.4 | FTO/$TiO_2$/NiO/$CH_3NH_3PbI_3$/Carbon | screen printing | [29] |
| commercial Carbon paste | 0.75 | 1.03 | 20.1 | 15.5 | FTO/$TiO_2$/mp-$TiO_2$/$CH_3NH_3PbI_3$/doped-TPDI/Carbon | doctor-blading | [30] |
| Carbon black/graphite (Carbon paste) | 0.76 | 0.917 | 21.36 | 14.9 | FTO/$TiO_2$/mp-$TiO_2$/$CH_3NH_3PbI_3$/$ZrO_2$/NiO/carbon black/graphite | blade coating | [31] |
| SWCNT doped graphite/Carbon black (Carbon paste) | 0.69 | 1.01 | 21.26 | 14.7 | FTO/compact $TiO_2$/$TiO_2$ and $Al_2O_3$ layer/$CH_3NH_3PbI_3$/Carbon SWCNTs | printing | [32] |
| Carbon paste (flake graphite/ graphite powder/ Carbon black/ $ZrO_2$ powder) | 0.474 | 0.97 | 23.5 | 10.8 | FTO/$TiO_2$/mp-$TiO_2$/$CH_3NH_3PbI_3$/Carbon/PDMS | doctor-blading | [33] |
| Carbon black/graphite (Carbon paste) | 0.572 | 0.457 | 19.63 | 5.13 | FTO/$TiO_2$/mp-$TiO_2$/$CH_3NH_3Sn_yPb_{1-y}I_{3-x}Cl_x$/ $Al_2O_3$/Carbon | screen-printing | [34] |
| Carbon paste | 0.46 | 0.843 | 14.19 | 5.56 | FTO/ZnO dense layer/$CH_3NH_3PbI_3$/(ZnO/$TiO_2$ NR layer)/$ZrO_2$ layer /Carbon layer | screen-printing | [35] |

| Electrode material | FF | $V_{OC}$ (V) | $J_{SC}$ (mA·cm$^{-2}$) | PCE (%) | Device structure | Method | Ref. |
|---|---|---|---|---|---|---|---|
| Carbon paste | 0.58 | 0.960 | 14.82 | 8.23 | FTO/ZnO dense layer/$CH_3NH_3PbI_3$/ZnO NR layer/$ZrO_2$ layer/ Carbon layer | screen-printing | [35] |
| soot of a burning candle | 0.42 | 0.82 | 12.30 | 4.24 | FTO/$TiO_2$/mp-$TiO_2$/$CH_3NH_3PbI_3$/ Spiro-OMeTAD/Carbon/FTO | using FTO glass as substrate to collect soot of a burning candle | [36] |
| Carbon black/graphite (Carbon paste) | 0.723 | 0.846 | 20.04 | 12.3 | FTO/$TiO_2$/mp-$TiO_2$/ $CH_3NH_3PbI_3$/$Al_2O_3$ /Carbon | screen-printing | [37] |
| low-temperature conductive Carbon paste (graphite flake, nano-graphite powder and Carbon black) | 0.44 | 0.85 | 18.3 | 6.88 | FTO/$TiO_2$/mp-$TiO_2$/$CH_3NH_3PbI_3$/Carbon | doctor-blading | [38] |
| Carbon black/graphite (Carbon paste) | 0.50 | 0.81 | 18.40 | 7.29 | FTO/$TiO_2$/mp-$TiO_2$/$CH_3NH_3PbI_3$/Carbon | doctor-blading | [39] |
| conductive Carbon ink | 0.60 | 0.89 | 21.43 | 11.44 | FTO/$TiO_2$/mp-$TiO_2$/$CH_3NH_3PbI_3$/Carbon | doctor-blading | [40] |
| commercial Carbon paste (Carbon black, graphite, and epoxy) | 0.65 | 1.04 | 21.27 | 14.38 | FTO/$TiO_2$/mp-$TiO_2$/$CH_3NH_3PbI_3$/Carbon | printing | [41] |
| Insulator cloth embedded in Carbon paste | 0.429 | 1.08 | 18.42 | 8.70 | FTO/$TiO_2$/mp-$TiO_2$/perovskite/Spiro-OMeTAD/Carbon | doctor-blading | [42] |

| Electrode material | FF | $V_{OC}$ (V) | $J_{SC}$ (mA·cm$^{-2}$) | PCE (%) | Device structure | Method | Ref. |
|---|---|---|---|---|---|---|---|
| Carbon cloth embedded in Carbon paste | 0.67 | 1.12 | 20.42 | 15.29 | FTO/$TiO_2$/mp-$TiO_2$/perovskite/Spiro-OMeTAD/Carbon | doctor-blading | [42] |
| commercial Carbon paste (Carbon black, graphite) | 0.67 | 1.0 | 21.83 | 14.58 | FTO/$TiO_2$/mp-$TiO_2$/$CH_3NH_3PbI_3$/Carbon | paint | [43] |
| Carbon paste | 0.75 | 0.893 | 22.43 | 15.0 | FTO/$TiO_2$/mp-$TiO_2$/$CH_3NH_3PbI_3$/$Al_2O_3$/Carbon | screen-printing | [44] |
| Carbon paste | 0.667 | 0.861 | 16.57 | 9.53 | FTO/$TiO_2$ compact layer/silver contact/$TiO_2$ mesoscopic layer/$CH_3NH_3PbI_3$/$ZrO_2$/Carbon | screen-printing | [45] |
| Carbon paste (Carbon black, graphite) | - | - | - | 11.07 | FTO/$TiO_2$/mp-$TiO_2$/$CH_3NH_3PbI_3$/mesoporous $ZrO_2$/Carbon | - | [46] |
| boron and phosphorus co-doped Carbon | 0.61 | 0.90 | 12.35 | 6.78 | FTO/$TiO_2$/mp-$TiO_2$/$CH_3NH_3PbI_3$/boron and phosphorus co-doped Carbon/Al foil | doctor-blading | [47] |
| Carbon black/graphite (Carbon paste) | 0.72 | 0.965 | 20.4 | 14.2 | FTO/$TiO_2$/mp-$TiO_2$/$CH_3NH_3PbI_3$/NiO nanosheets/Carbon | screen-printing | [48] |
| Carbon | 0.59 | 0.857 | 20.79 | 10.53 | FTO/$TiO_2$ dense layer/W-doped mesoporous $TiO_2$/ $CH_3NH_3PbI_3$/Carbon | blade coating | [49] |
| commercial Carbon paste | 0.74 | 1.05 | 20.8 | 16.1 | FTO/$TiO_2$/mp-$TiO_2$/$CH_3NH_3PbI_3$/CuPc nanorods/Carbon | doctor-blading | [50] |
| Carbon slurry | 0.67 | 0.867 | 22.93 | 13.41 | FTO/$TiO_2$/mp-$TiO_2$/$ZrO_2$/perovskite/Carbon | screen-printing | [51] |
| graphite/Carbon black (Carbon paste) | 0.78 | 0.92 | 19.21 | 13.89 | FTO/$TiO_2$/mp-$TiO_2$/$ZrO_2$/perovskite/Carbon | screen printing | [52] |
| Carbon paste | 0.77 | 0.94 | 21.45 | 15.60 | FTO/$TiO_2$/mp-$TiO_2$/$ZrO_2$/perovskite/Carbon | screen printing | [53] |

| Electrode material | FF | $V_{OC}$ (V) | $J_{SC}$ ($mA\cdot cm^{-2}$) | PCE (%) | Device structure | Method | Ref. |
|---|---|---|---|---|---|---|---|
| commercial Carbon paste | 0.72 | 1.35 | 8.35 | 8.09 | $FTO/TiO_2/mp\text{-}TiO_2/CH_3NH_3PbBr_3/Carbon$ | printing | [54] |
| commercial Carbon paste (Carbon black/ graphite) | 0.68 | 1.29 | 5.7 | 5.0 | $FTO/TiO_2/mp\text{-}TiO_2/CsPbI_3/Carbon$ | painting | [55] |
| graphite | 0.644 | 0.803 | 20.1 | 10.4 | $FTO/TiO_2/mp\text{-}TiO_2$/Perovskite/Carbon | doctor-blading | [56] |
| commercial Carbon paste | 0.55 | 0.97 | 19.1 | 10.19 | $FTO/TiO_2/mp\text{-}TiO_2$/Perovskite/Carbon | printing | [57] |
| Carbon paste | 0.65 | 1.04 | 21.27 | 14.38 | FTO/ $TiO_2/mp\text{-}TiO_2$/Perovskite/Carbon | - | [58] |
| Carbon paste | 0.68 | 0.88 | 15.10 | 9.10 | $FTO/TiO_2/mp\text{-}TiO_2/ZrO_2$/Perovskite($MAPbI_3$)/Carbon | screen printing | [59] |
| Carbon paste | 0.74 | 1.00 | 19.31 | 14.35 | $FTO/TiO_2/mp\text{-}TiO_2/ZrO_2$/Perovskite ($MAPbI_3$ ▪xGuCl)/Carbon | screen printing | [59] |
| Carbon black/graphite (Carbon paste) | 0.69 | 1.04 | 15.37 | 11.03 | $FTO/TiO_2/mp\text{-}TiO_2/Al_2O_3$/Perovskite/Carbon | screen printing | [60] |
| Carbon paste | 0.76 | 0.957 | 18.15 | 13.24 | $FTO/TiO_2/mp\text{-}TiO_2/CH_3NH_3PbI_{(3-x)}(BF_4)_x/ZrO_2/Carbon$ | screen-printing | [61] |
| Candle soot | 0.72 | 0.90 | 17.00 | 11.02 | $FTO/TiO_2/mp\text{-}TiO_2/CH_3NH_3PbI_3$/Carbon /FTO | rolling transfer | [62] |
| Carbon paste | 0.66 | 0.90 | 20.7 | 12.30 | $FTO/TiO_2/mp\text{-}TiO_2$/ porous $Al_2O_3/CH_3NH_3PbI_3$/Carbon | screen printing | [63] |
| Carbon paste | 0.68 | 0.78 | 15.1 | 8.0 | $FTO/TiO_2/mp\text{-}TiO_2$/porous $Al_2O_3$/Perovskite/Carbon | doctor-blading | [64] |
| Carbon paste | 0.53 | 1.00 | 22.67 | 12.02 | $FTO/TiO_2/mp\text{-}TiO_2/CH_3NH_3PbI_3/Carbon$ | doctor-blading | [65] |

| **Electrode material** | **FF** | **$V_{OC}$ (V)** | **$J_{SC}$ ($mA \cdot cm^{-2}$)** | **PCE (%)** | **Device structure** | **Method** | **Ref.** |
|---|---|---|---|---|---|---|---|
| laminated Carbon nanotube networks | 0.51 | 0.88 | 15.46 | 6.87 | FTO/$TiO_2$/mp-$TiO_2$/$CH_3NH_3PbI_3$/CNT film | transfer | [66] |
| Carbon nanotubes | 0.68 | 0.99 | 14.36 | 8.31 | Ti foil/$TiO_2$ NTs/ $CH_3NH_3PbI_3$/spiro-OMeTAD/Carbon nanotubes | transfer | [67] |
| single-walled Carbon nanotube (SWCNT) | 0.61 | 1.1 | 20.3 | 13.6 | FTO/$TiO_2$/mp-$TiO_2$/ $(FAPbI_3)_{0.85}(MAPbBr_3)_{0.15}$/ Spiro-OMeTAD/SWCNT | press transfer | [68] |
| cross-stacked superaligned Carbon nanotube (CSCNT) sheets with 25-layer CNTs | 0.53 | 0.878 | 14.43 | 6.81 | FTO/$TiO_2$/mp-$TiO_2$/$CH_3NH_3PbI_3$/$Al_2O_3$/CSCNTs/PMMA | transfer | [69] |
| cross-stacked superaligned Carbon nanotube (CSCNT) sheets with 50-layer CNTs | 0.68 | 0.849 | 14.91 | 8.60 | FTO/$TiO_2$/mp-$TiO_2$/$CH_3NH_3PbI_3$/$Al_2O_3$/CSCNTs/PMMA | transfer | [69] |
| iodine doping cross-stacked superaligned Carbon nanotube (CSCNT) sheets with 50-layer CNTs | 0.71 | 0.853 | 17.22 | 10.54 | FTO/$TiO_2$/mp-$TiO_2$/$CH_3NH_3PbI_3$/$Al_2O_3$/CSCNTs/PMMA | transfer | [69] |
| cross-stacked superaligned Carbon nanotube (CSCNT) sheets with 75-layer CNTs | 0.64 | 0.823 | 15.81 | 8.35 | FTO/$TiO_2$/mp-$TiO_2$/$CH_3NH_3PbI_3$/$Al_2O_3$/CSCNTs/PMMA | transfer | [69] |
| thermally treated cross-stacked superaligned Carbon nanotube (CSCNT) sheets with 50-layer CNTs | 0.69 | 0.840 | 16.21 | 9.37 | FTO/$TiO_2$/mp-$TiO_2$/$CH_3NH_3PbI_3$/$Al_2O_3$/CSCNTs/PMMA | transfer | [69] |

| **Electrode material** | **FF** | **$V_{OC}$ (V)** | **$J_{SC}$ ($mA \cdot cm^{-2}$)** | **PCE (%)** | **Device structure** | **Method** | **Ref.** |
|---|---|---|---|---|---|---|---|
| multi-walled Carbon nanotubes | 0.59 | 0.926 | 21.3 | 11.6 | FTO/$TiO_2$ dense layer/($TiO_2$/$SiO_2$ scaffold layer)/$CH_3NH_3PbI_3$/ Carbon | doctor-blading | [70] |
| Boron Doped Multi-walled Carbon Nanotubes | 0.77 | 0.92 | - | 15.23 | FTO/$TiO_2$/mp-$TiO_2$/$CH_3NH_3PbI_3$/$Al_2O_3$/B-doped MWNT | drop-casting | [71] |
| multi-walled Carbon nanotube (MWCNT) | 0.80 | 0.88 | 18.00 | 12.67 | FTO/$TiO_2$/$CH_3NH_3PbI_3$/MWCNT | drop cast | [72] |
| NiO/SWNCT | 0.64 | 0.945 | 20.7 | 12.7 | FTO/$TiO_2$/mp-$TiO_2$/$Al_2O_3$/NiO/SWNCT | screen printing | [73] |
| graphene | 0.67 | 0.96 | 19.17 | 12.37 | FTO/compact-$TiO_2$/$CH_3NH_3PbI_{3-x}Cl_x$/Spiro-OMeTAD/PEDOT: PSS/graphene/PMMA/PDMS | laminate | [74] |
| three-dimensional honeycomb like structured graphene (3DHG) | 0.63 | 0.89 | 18.11 | 10.06 | FTO/$TiO_2$/($CH_3NH_3PbI_3$/3DHG)/ 3DHG | doctor-blading | [75] |
| Single-layered graphene(SG) | 0.54 | 0.878 | 14.2 | 6.7 | FTO/$TiO_2$/Perovskite bilayer/SG | spin-coating | [76] |
| Multilayered graphene(MG) | 0.73 | 0.943 | 16.7 | 11.5 | FTO/$TiO_2$/Perovskite bilayer/MG | spin-coating | [76] |

## 2. Carbon Materials Based Counter Electrodes

Carbon materials have many merits including abundant sources, low-cost, high electrical conductivity, chemical stability, diversity, modifiability[77]. The appropriate 5.0eV work function near to Au (5.1eV) makes carbon an ideal candidate for PVSCs' back electrode. According to the different bonded patterns of carbon atoms and different dimensions of crystalline structures, there are several allotropes with very different chemical and physical properties in the elemental carbon family, such as diamond, graphite, graphene, graphdiyne, C60, carbon nanotube, and amorphous carbon, as shown in Fig. 2.[78]

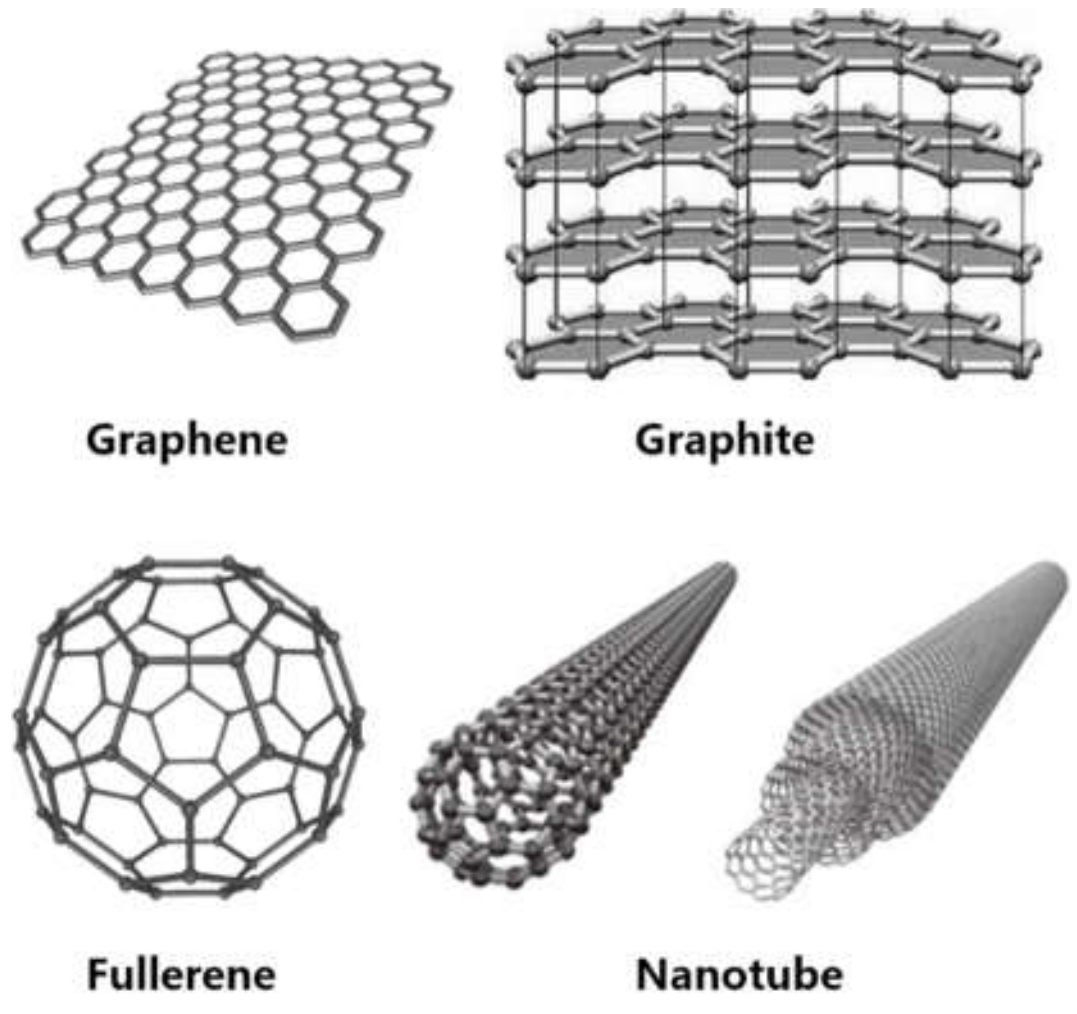

**Fig. 2.** Structure of various carbon materials.

### 2.1. Carbon paste

Carbon paste was mainly prepared by graphite and carbon black and was often formed on the top of hole-conductor-free PVSCs. Graphite is soft, soapy, hexagonal layered crystal with some particular characteristics such as high-temperature resistance, excellent electrical conductivity and thermal conductivity, lubricity, anti-corrosion, plasticity and thermal shock resistance. Carbon black is light, loose, and extremely fine black powder produced by incomplete combustion or thermal decomposition of carbon aceous matter under insufficient air condition. Carbon black is one kind of amorphous carbon. The so-called 'amorphous carbon ' isn't indicating that the shapes of carbon black are confused, but it means the inner turbostratic structure of stacking graphite layers.[78] There are mainly two types of device configurations in carbon CE based PVSCs. One is called the "monolithic structure" which was previously reported in DSSCs featuring the loading of perovskite in the last step [79] (Fig. 3a), and the other one is "conventional structure" (Fig. 3b) which is fabricated by a 'layer-by-layer' process and a condensed carbon layer atop. In this section, we will discuss carbon paste based CEs used in these two device configurations.

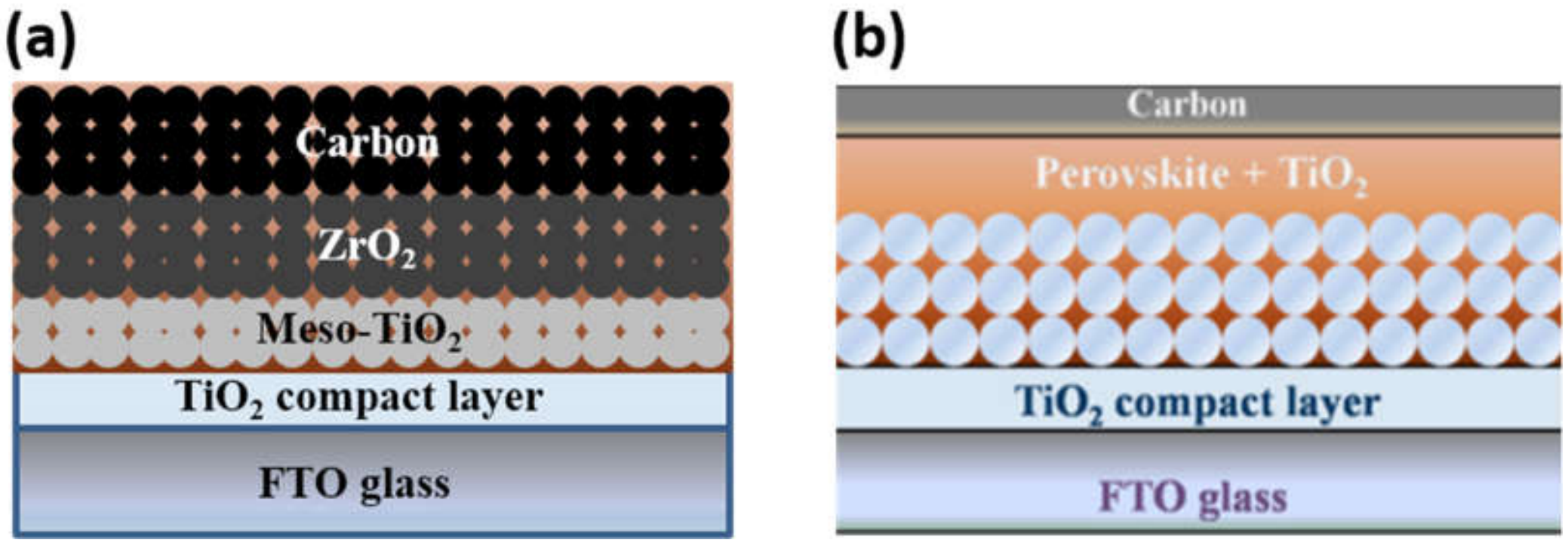

**Fig. 3.** A schematic structure of device configurations in carbon CE based PVSCs. (a) Monolithic structure; (b) conventional structure. Reprinted from ref. [21]

**2.1.1.** Monolithic structure

In 2013, Hongwei Han's group established a hole-conductor-free $CH_3NH_3PbI_3$ perovskite/$TiO_2$ heterojunction solar cell which first introduced carbon black/ spheroidal graphite as a counter electrode to substitute noble metallic electrode, and an impressive PCE value of exceeding 6.64% was obtained. This carbon black/graphite back electrode was screen-printed on the top of $TiO_2$ nanocrystalline layer, then the $CH_3NH_3PbI_3$ precursor was dipped on top of the mesoscopic carbon layer, and finally formed a device configuration of FTO glass substrate/$TiO_2$ dense layer/$TiO_2$ mesoscopic layer/$ZrO_2$ space layer/carbon layer, which is named "monolithic structure" [25,79]. The device was fabricated by a simple process like screen printing and shows a long-term stability in the dark after 840 hours. This work initiated the research to low-cost carbon back electrode based PVSCs.[18] Later, they did a series of work on this carbon -based electrode PVSCs.

In 2014, they introduced a mixed-cation perovskite $(5\text{-AVA})_x(MA)_{1-x}PbI_3$ to replace $MAPbI_3$ in almost the same triple-layer structure of $TiO_2/ZrO_2$/Carbon, as shown in Fig. 4. The 5-AVA molecules formed linear hydrogen-bonded chains between their COOH and $NH_3^+$ groups and $I^-$ ions from the $PbI_6$ octahedra, acting as a templating agent to create mixed-cation perovskite $(5\text{-AVA})_x(MA)_{1-x}PbI_3$ crystals with lower defect concentrations and better pore filling as well as more complete contact with the $TiO_2$ scaffold, resulting in a longer exciton lifetime and a higher quantum yield for photoinduced charge separation as compared to $MAPbI_3$. The cell achieved a certified PCE of 12.8 % and showed long-time stability(>1000h) in ambient air under full sunlight.[23] Mixed anion strategy that makes $BF^{4-}$ substitute for fractional $I^-$ was also employed, showing a PCE enhancement from 10.54% to 13.24% due to more efficient charge transport and suppressed recombination.[61]

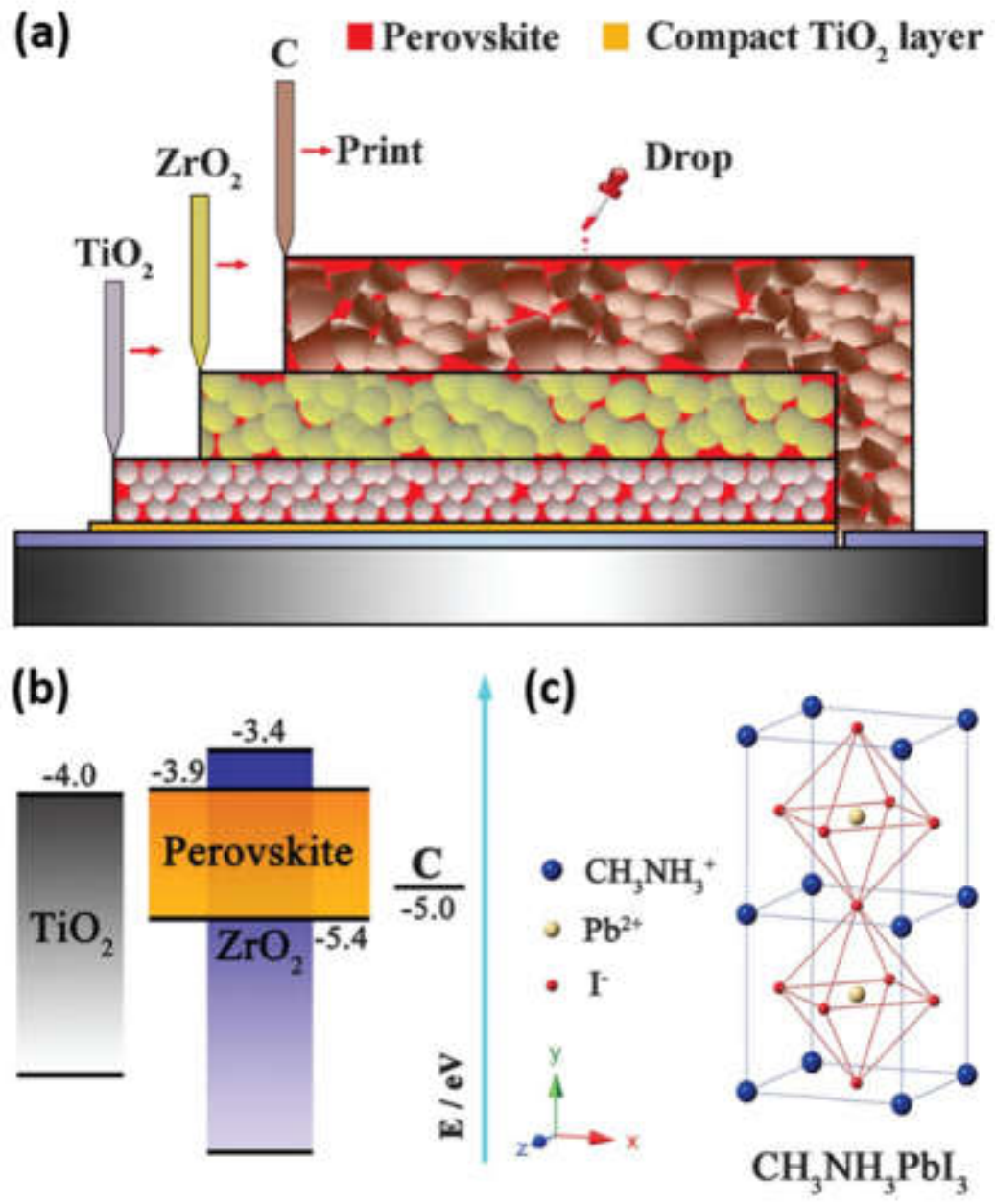

**Fig. 4.** (a) The crystal structure of $CH_3NH_3PbI_3$ perovskite and the corresponding energy levels of $TiO_2$, $CH_3NH_3PbI_3$, and carbon. (b) A schematic structure of the carbon-based monolithic device. Reprinted from ref.[18]

It is known that device performance depends heavily on the quality of the perovskite layer such as morphology, crystallinity, and density of defects. In the case of mesoporous device structures, higher pore filling of perovskites phase is crucial to achieving high-performance monolithic PVSCs, which generally have thick mesoporous films over 10$\mu$m. Thus, they further studied the solvent effect of the perovskite and found that optimized DMF/DMSO (0.93:0.07, v/v) showed a better wettability because of a decent compromise of polarity and viscosity of the solvent. The as-prepared PVSCs hold better perovskite infiltration, the light harvesting ability and charge separation of the perovskite layer were improved, and then the cell devices displayed a PCE of 13.89%.[52] They also employed a versatile additive of guanidinium chloride to improve the morphology of $CH_3NH_3PbI_3$ perovskite absorber. An improved contact between the perovskite and carbon layer suppresses the recombination reaction in the device, and the open-circuit voltage of the device is significantly enhanced from 0.88 V to 1.02 V, resulting in a PCE up to 14.35%.[59] Besides, $NH_4Cl$ was added to assist the crystallization of perovskite. A synergy effect of $NH_4Cl$ and moisture on $CH_3NH_3PbI_3$ crystallization provides a facile way to accomplish complete pore-filling of perovskite absorbers in the mesoporous triple-layer scaffold in printable PVSCs. Correspondingly, the devices achieve a champion efficiency of 15.60% with $J_{SC}$ = 21.45 mA cm$^{-2}$, $V_{OC}$ = 0.94 V and FF = 0.77 and a lifetime of over 130 days in ambient air with relative humidity of 35%.[53] More parameters including loading amount and morphology of $MAPbI_3$, the temperature of the substrate and precursor relative with the quality of perovskite were optimized in the two-step method fabrication process.[26]

Except for the quality of perovskite itself, they investigated the size effect of bottom

$TiO_2$ nanoparticles on their printed PVSCs. The size of $TiO_2$ particles affects the infiltration of the precursor and the contact between the perovskite crystal and $TiO_2$, while influences the charge transfer at the perovskite/$TiO_2$ interface. The optimized diameter of 25 nm $TiO_2$ nanoparticles exhibits the best PCE of 13.41%.[51]

The $ZrO_2$ film is also very important, who separates the $TiO_2$ film and carbon film, avoiding any direct contact while influences light harvesting and the transfer of photo-generated charge. They found that appropriate thickness of the mesoporous $ZrO_2$ film was around 1μm, as a result of a compromise between needs of light harvesting by $MAPbI_3$ confined in spacer layer, the width of the built-in electric field, and film quality of spacer layer.[26]

Also, as a pivotal component, graphite decides the porosity and conductivity of back carbon electrode. Different sizes of graphite will cause differences in the filling of $PbI_2$ and $CH_3NH_3PbI_3$ precursors and also the square resistance of carbon counter electrode, which will finally affect the performance of devices. So, they investigated the previous topic, and they found that the 8 mm graphite-based carbon back electrode has a more significant average pore size, a smaller square resistance and hence a higher PCE exceeding 11% compared with others. The square resistance of 9 mm thick carbon CEs employing graphite with different sizes was showed in Table 2. [80]

Table 2. The square resistance of 9 um thick carbon CEs employing graphite with different sizes. Reproduced from ref. [80].

| | 8μm | 6 μm | 3μm | 500 nm | 40 nm |
|---|---|---|---|---|---|
| $R_{sq}(\Omega)$ | 11.47 | 61.21 | 132.26 | 56.68 | 63.01 |

Besides Han's work, to realize more mechanization of the fabrication devices process, Hashmi et al.[45] replaced manual infiltration of perovskite precursor step used in Han's works with inkjet printing in the same device structure of PVSCs (Fig. 5. (a), (b)), and the fabricated devices showed a maximum PCE of 9.53% with high stability when subjected to continuous light soaking stability test over a period of 1046 h without any encapsulation. This work demonstrated the capability, and potential of inkjet infiltration of perovskite precursor ink for porous triple layered HTM free printed PVSCs and provides an opportunity to fabricate the large-area devices in future.

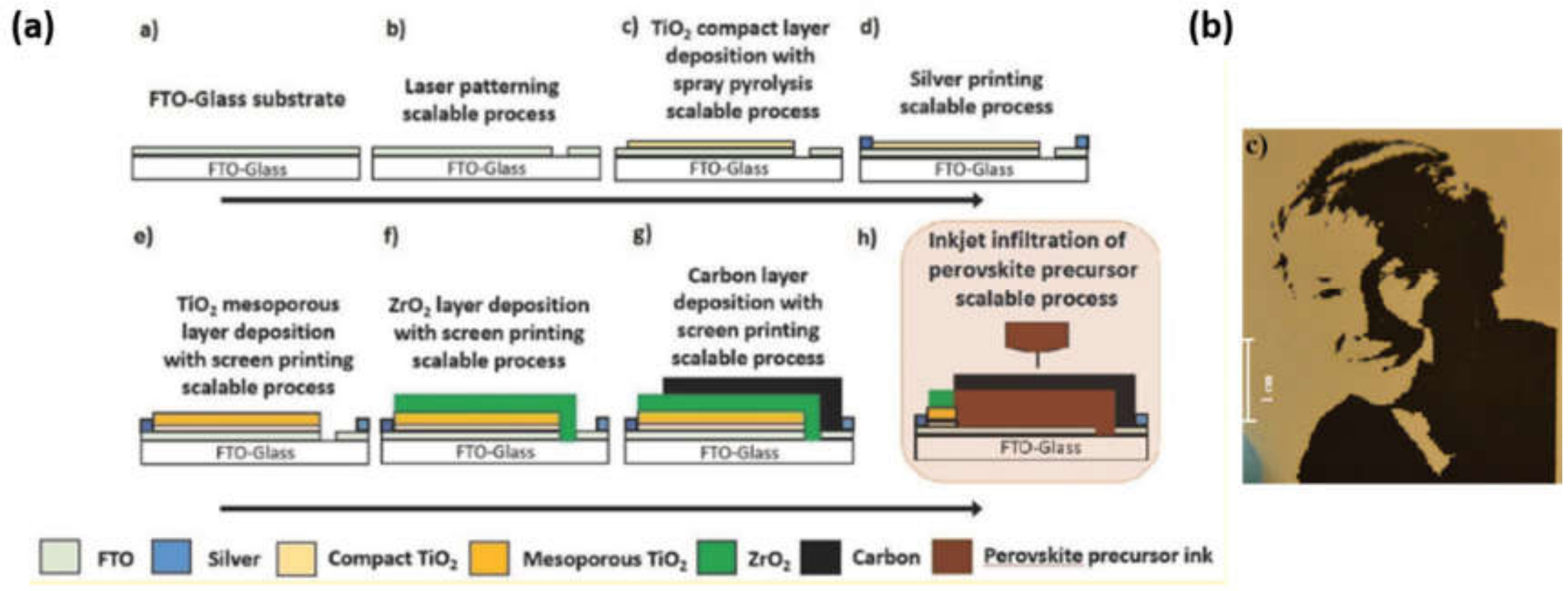

**Fig. 5.** (a) Illustration of the scalable process to be employed for the production of HTM free carbon

based PVSCs. a) Fluorine-doped tin oxide (FTO) coated glass substrate, b) laser patterning of the FTO layer, c) spray pyrolysis for compact $TiO_2$ layer deposition, d) screen printing of silver contacts, e) screen printing of a mesoporous $TiO_2$ layer, f) screen printing of an insulating $ZrO_2$ layer, g) screen printing of a porous carbon composite layer, h) inkjet of a controlled volume of perovskite precursor solution, which is targeted for this work. Note: Only scalable processes are highlighted here. The sintering steps consecutive to screen printing are not represented for better clarity. (b) Inkjet printing of bitmap version of a digital image as examples of accurate 2D printing, image reproduced with permission from Alain Herzog/EPFL. Adapted from ref.[18]

To reduce the interfacial charge recombination, Mingkui Wang's group developed a quadruple layer of $TiO_2/Al_2O_3$/NiO/Carbon as scaffold infiltrated with $MAPbI_3$, as shown in Fig. 6. The insulating $Al_2O_3$ act as a spacer layer, while the n-type $TiO_2$ and p-type NiO act as selective contacts for the extraction of electrons and holes, respectively. The $TiO_2/Al_2O_3$/NiO/Carbon structured device showed a short-circuit current density ($J_{SC}$) of 21.62 mA/cm$^2$, a fill factor (FF) of 0.76, an open-circuit voltage ($V_{OC}$) of 915 mV, achieving a PCE of 15.03%. For comparison, the device with a configuration of $TiO_2/Al_2O_3$/Carbon ($MAPbI_3$) was also fabricated, showing a lower PCE performance of 11.2% ($J_{SC}$ = 17.59 mA/cm$^2$, FF = 0.71, and $V_{OC}$ = 896 mV). Incident photon to current conversion efficiency (IPCE) curves of these two kinds of devices were measured, the $TiO_2/Al_2O_3$/NiO/Carbon ($MAPbI_3$) device reaches a higher value than the $TiO_2/Al_2O_3$/Carbon ($MAPbI_3$) device in the wavelength range of 350–750 nm. The augmented $\Delta$ IPCE in the blue region corresponds to carriers generated by high-energy photons in the front section close to the FTO side, indicating that these charges would be efficiently collected when the NiO layer was inserted, most probably due to the direct separation of the electron/hole pair at this interface without having to diffuse through the perovskite itself. Although no organic hole-transport layer was employed to collect holes, a high FF of 0.76 is still observed for the $TiO_2/Al_2O_3$/NiO/Carbon ($MAPbI_3$) device. The high FF and IPCE values indicate that the charge transport and collection are very efficient in this device architecture.[25]

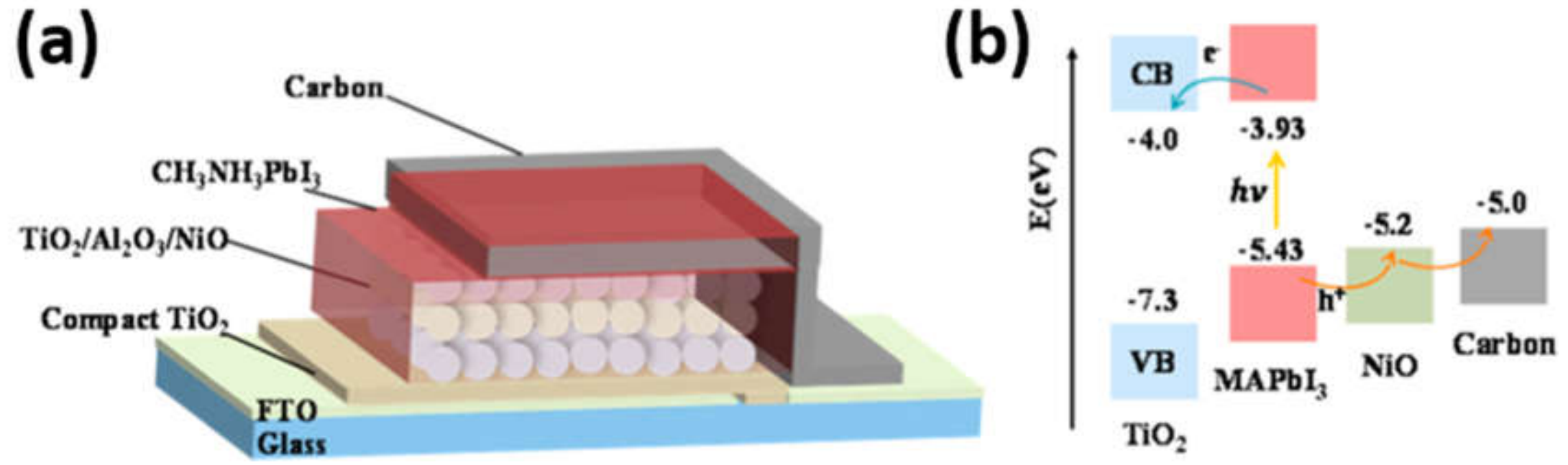

**Fig. 6.** Optimized device structure. (a) Schematic representation of the $TiO_2/Al_2O_3$/NiO/carbon ($MAPbI_3$) based device. (b) Energy band diagram of the fabricated device configuration. Reprinted from ref.[25]

The perovskites adopted in works mentioned above are all organic-inorganic hybrid perovskites. Recently, inorganic perovskites (e.g. $CsPbI_3$, $CsPbBr_3$) attracted extensive attention due to higher stability. [81-83] A printed HTM-free PVSCs with all the components made of inorganic materials (porous metal oxide, inorganic perovskite, and carbon) with monolithic configuration showed increased thermal stability. Li et al.[84] fabricated lead-free inorganic perovskites $CsSnX_3$-based PVSCs with the structure of $FTO/TiO_2/Al_2O_3/CsSnX_3/C$, the as-prepared all-inorganic devices showing superior thermal stability up to 200 °C, with an average PCE of 3.0% and long-term stability for 77 days without efficiency-loss.

In all the works above, the monolithic PVSCs were usually fabricated by inexpensive and straightforward screen-printing method, of which the structure avoids costly and unstable organic HTMs and noble metal electrode like Au. These HTM-free carbon-based PVSCs boost the promise of PVSC to commercialization. The monolithic PVSCs have the advantages of ambient condition process, efficacious use of precursor, thick perovskite layer for higher stability and better contact between perovskite and CE. Meanwhile, the thick metal oxide layers bring disadvantages like the high-temperature process and large series resistance, etc.

**2.1.2.** Conventional structure

Regarding the carbon paste mentioned above, high-temperature sintering above 400 °C is necessary, which will hamper the large-scale commercialization of carbon-based PVSCs. So, it needs to develop the low-temperature processed CE. Apart from the lower energy consumption, low-temperature (<150 °C) permits not only to make electrodes straight on top of the perovskites but also prepare flexible devices. For example, it is possible to print carbon pastes directly on perovskite layers like deposition of Au on top of the perovskites or HTM (e.g., Spiro-OMeTAD) in a conventional PVSC. In 2014, Licheng Sun's group first reported a low-temperature-processed commercial carbon paste (100 °C) by doctor blading technology. They deposited the carbon electrode on the two-step method prepared perovskites. Under optimized conditions, the hole-conductor-free perovskite/$TiO_2$ heterojunction solar cells obtained an impressive PCE value of 8.31% and furthermore exhibited excellent stability over 800 h.[21] The structure of perovskite would be destroyed by some solvents in commercial carbon paste which is directly processed on the top of the perovskite film. Thus, Tingli Ma's Group[85] modified the commercial conductive carbon paste with only chlorobenzene as a solvent and obtained an improved PCE of 9.08%.

At almost the same time, Yang's group reported clamping solar cells using candle soot for hole-extraction. In the fabrication process, the perovskite film is fabricated by depositing a $PbI_2$ precursor layer, immediately followed by the rolling transfer of candle soot, and finally subjecting the sample to a $CH_3NH_3I$ bath treatment, as shown in Fig. 7. (a-c) Notably, the hole-extraction rate of candle soot is even comparable with that of traditional spiro-OMeTAD, which was demonstrated by the femtosecond time-resolved photoluminescence (fs-TRPL). The best devices achieve a PCE of 11.02%.[62] Later, they developed an inkjet printing technology to fabricate carbon electrodes in planar $TiO_2$ based PVSCs. The fabrication process was diagrammatized in Fig. 7. (d). By employing the carbon/$CH_3NH_3I$ ink formulation to transform $PbI_2$ in situ to $CH_3NH_3PbI_3$, a reinforced interpenetrating interface between the $CH_3NH_3PbI_3$ and carbon electrodes was formed in comparison with that using pure carbon ink. With minimal charge recombination, a reasonably high PCE of 11.60% was achieved.[19] This type of device processing combined the procedures from monolithic and conventional devices. Their strategy of in-situ preparation of perovskite builds an excellent contact between perovskite and CEs, which could avoid the possible adverse effect of carbon paste solvent. However, this method exists apparent defects where their $J_{SC}$ is as low as only 17 mA·cm$^{-2}$, which might be ascribed to the thin perovskite layer produced by the two-step transformation reaction ($PbI_2$ layer to perovskite) and always along with unreacted $PbI_2$.

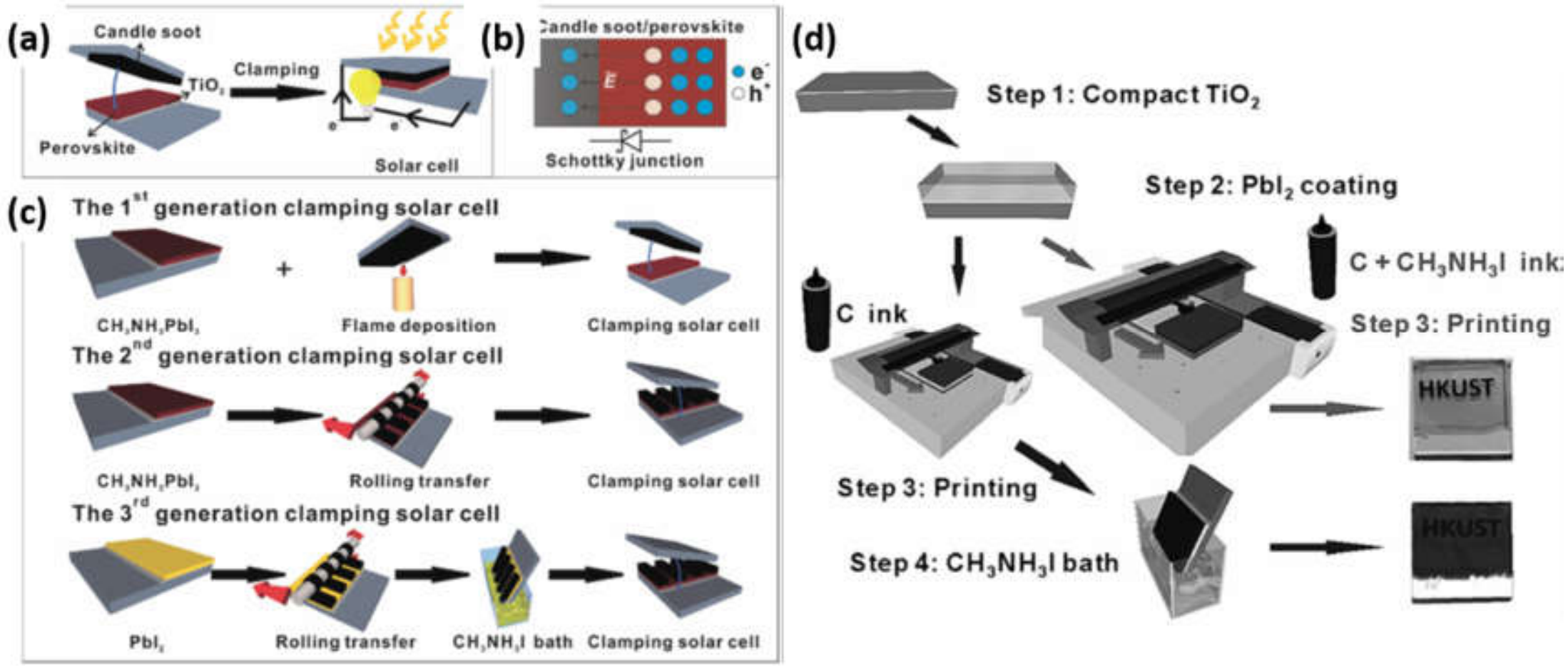

**Fig. 7.** (a), (b), (c) Development of a candle soot/perovskite clamping solar cell. Reprinted from ref. [62] (d) Fabrication process flow for the instant inkjet printing of the C/$CH_3NH_3PbI_3$ planar perovskite solar cells. For comparison, a different strategy was used to convert $PbI_2$ into $CH_3NH_3PbI_3$ using a separate step 3 and step 4. Reprinted from ref. [19]

Then, Yang's group move efforts towards the fabrication of high-quality perovskites by painting carbon pastes on the top of pre-prepared perovskites. They developed electrodeposition technique to synthesize the perovskite layer. The electrodeposited perovskite layer was printed with the carbon paint as back contact. The corresponding devices achieved a PCE of 10.19%, and show an encouraging PCE in a larger cell area up to 4 cm$^2$.[57] Although the efficiency of painted carbon-based-PVSCs has come up to over 10%, it is still low compared with the traditional PVSCs with Au or Ag CEs. The primary limiting factor of the efficiency of painted-carbon-based PVSC lies in the poor contact at perovskite/carbon interface due to the lamella feature of graphite layer and ease of producing interspace between the graphite flakes and perovskites.[20]

To improve contact at perovskite/carbon interface, Yang et al. designed a solvent engineering strategy based on two-step sequential method which is beneficial to an excellent pore-filling perovskite layer in a relatively thick mesoscopic $TiO_2$ scaffold. The solvent for $CH_3NH_3I$ solution at the second step is a mixed solvent of IPA/Cyclohexane. This mixed solvent results in a high-quality perovskite layer with smooth surface and compact capping layer which could enhance the contact with CE for a majority of graphite flake in carbon paste. Finally, they obtained the highest PCE of 14.38% with an FF of 0.65 and a PCE of 10% for the 1 $cm^2$ area device.[41] Moreover, through colloidal engineering to form ultra-even perovskite layer, they boosted the efficiency of the carbon-based PVSCs to 14.58% [43]

In the presence of high-quality perovskite layer, Sun et al. introduced CuPc nanorods as hole-transporting materials to improve the interface contact. The device displayed an impressive PCE of 16.1%, which is comparable to the corresponding devices with the doped spiro-OMeTAD as HTMs and noble metal Au as back electrode. CuPc nanorods induce an intimate, large interfacial-area contact with the $MAPbI_3$ crystal grains and carbon counter electrode, therefore leading to a significant enhancement in hole- extraction, and a decrease of charge recombination.[50] (Fig. 8.) However, the CuPc nanorods were prepared by thermal deposition that needs high-vacuum condition.

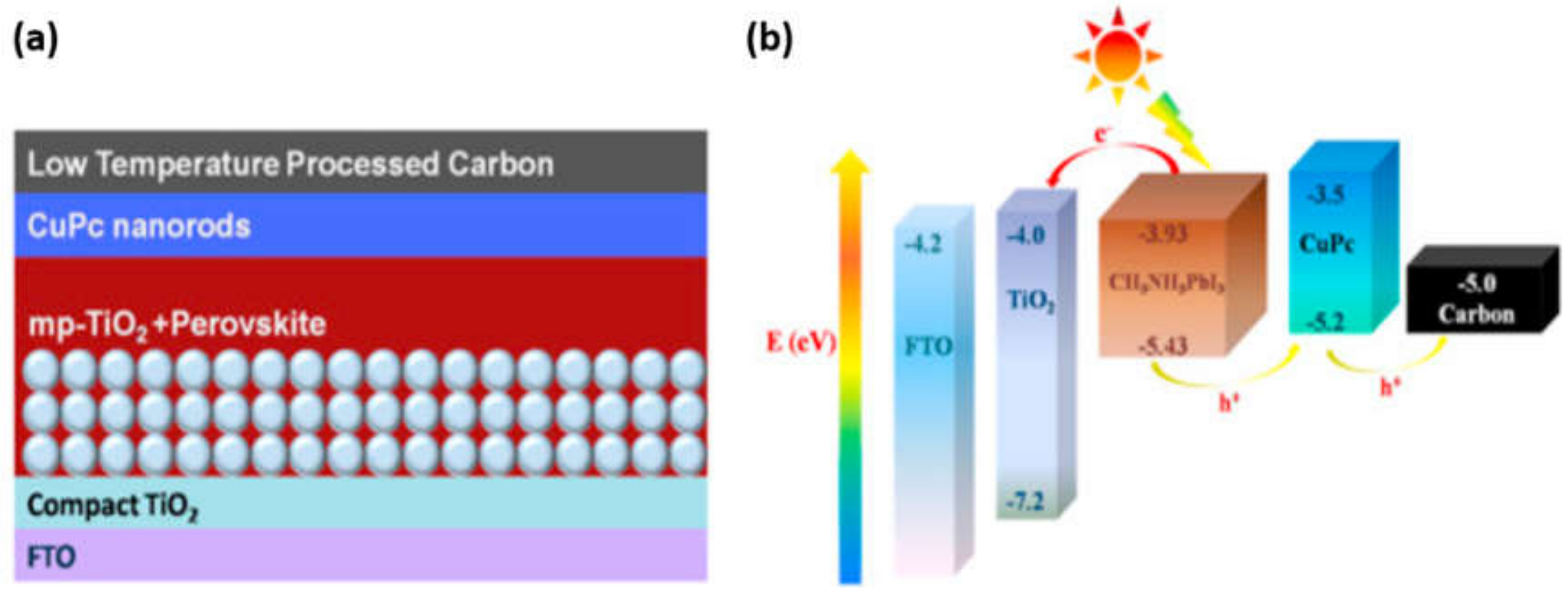

**Fig. 8.** Device structure and energy-level diagram: (A) Schematics of the whole device: FTO glass/compact $TiO_2$ layer/mesoporous $TiO_2$/$CH_3NH_3PbI_3$ capping layer/CuPc/carbon and the corresponding (B) energy level diagram. Fabrication process flow for the instant inkjet printing of the C/$CH_3NH_3PbI_3$ planar perovskite solar cells. For comparison, a different strategy was used to convert $PbI_2$ into $CH_3NH_3PbI_3$ using a separate step 3 and step 4. Reprinted from ref. [50]

**2.1.3.** Device engineering on carbon CEs

To enhance the contact of CE and perovskite layer for the favorable charge transport, Qingbo Meng's group developed two-step processed carbon CEs with different graphite flake sizes and contents of carbon black. Firstly, viscous carbon pastes were first coated on the top of the as-prepared devices. Then, a piece of graphite paper was pressed on the carbon paste to as an extraction electrode, as shown in Fig. 9. The electrochemical impedance spectroscopy (EIS) of the different cells with carbon CEs was measured to study the charge transport properties. With the lowest charge transfer resistance ($R_{CT}$) level, the graphite/carbon black composite based cell exhibited a best PCE of 10.2% and a highest FF of 0.572, which suggested that the placid undulation at the surface of smaller graphite flakes, and together with the carbon black particles filling in the interspace create more contact sites with the perovskite layer than other cells without doping with carbon black.[20]

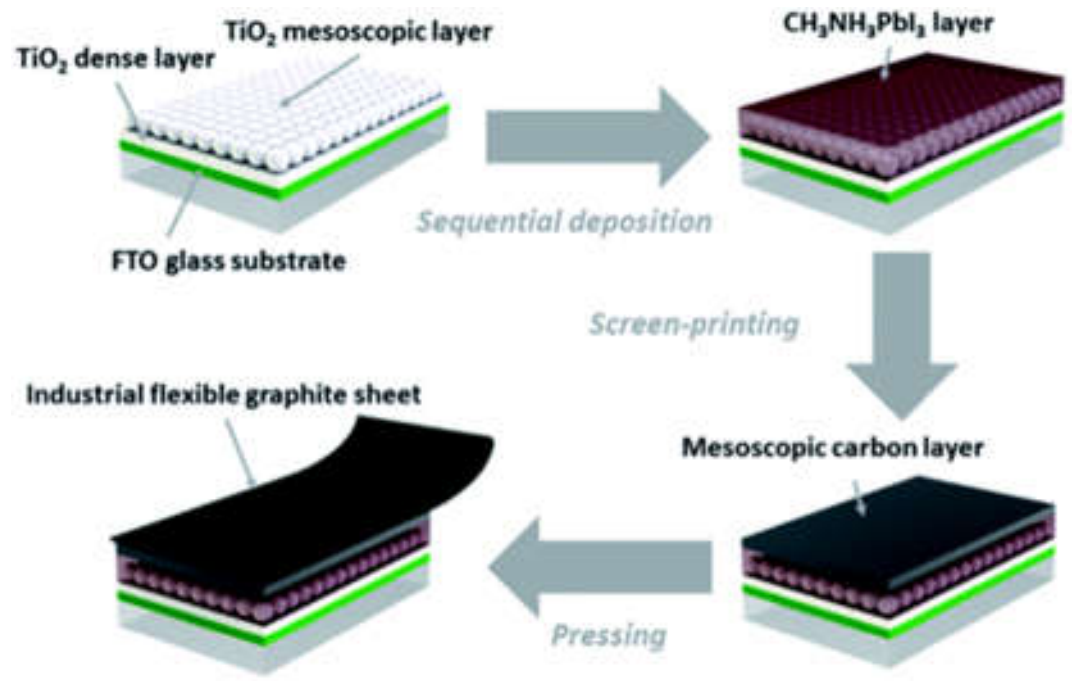

**Fig. 9.** Scheme of the fabrication process of all-carbon flexible CE based HTM-free perovskite solar cells. Reprinted from ref.[20]

Meng et al. also fabricated a free-standing flexible carbon electrode containing 20%wt polyvinyl acetates (PVAc) in the carbon film to avoid the adverse effect of solvents in carbon pastes. Carbon paste was coated on Teflon film by doctor-blading to form the carbon film after the solvent evaporation. Due to the thermoplastic PVAc, the carbon films can be directly hot-pressed onto the surface of perovskite at 85°C. With the optimized carbon film components (graphite flakes and carbon black particles with a weight ratio of 3:1 and hot-press pressure (0.25 MPa), champion PCE of 13.53% with an average PCE of 12.03% have been achieved. Compared to PVSCs with Au electrode, this low-cost and stable carbon-based one with low-temperature-processed and flexible carbon counter electrode is promising.[27]

A low-cost carbon cloth embedded in carbon paste as CE were studied by Hagfeldt et al., as shown in Fig. 10(a). When compared to an insulating fiber embedded in carbon paste, the carbon cloth composite obtained an enormous increase in FF. These results showed that the carbon cloth is an essential for achieving high-efficiency carbon devices. By a press-transfer process of carbon microfibers in carbon paste, high $J_{SC}$ of 20.4 mA cm$^{-2}$ and high $V_{OC}$ of 1.12 V were achieved. Moreover, a champion stabilized PCE of 14.8% was achieved.[42]

Additionally, Liao et al. reported a low-temperature carbon CE based HTM-free mesoscopic PVSCs and encapsulated them by PDMS, as shown in Fig. 10(b). The introduction of PDMS can condense the mesoscopic carbon layer and improve the $MAPbI_3$/carbon interface condition during the solidification of PDMS. Finally, an optimal PCE of 10.8% was achieved. The results exhibit a remarkable 54% enhancement over those without encapsulation (a PCE of 6.88%). The improved performance can be attributed to efficient charge transfer process and slower charge recombination between interfaces. Also, the PDMS layer can isolate moisture in the air, avoiding the degradation of the perovskite. The encapsulated cells show dramatic stability over 3000 hours.[33]

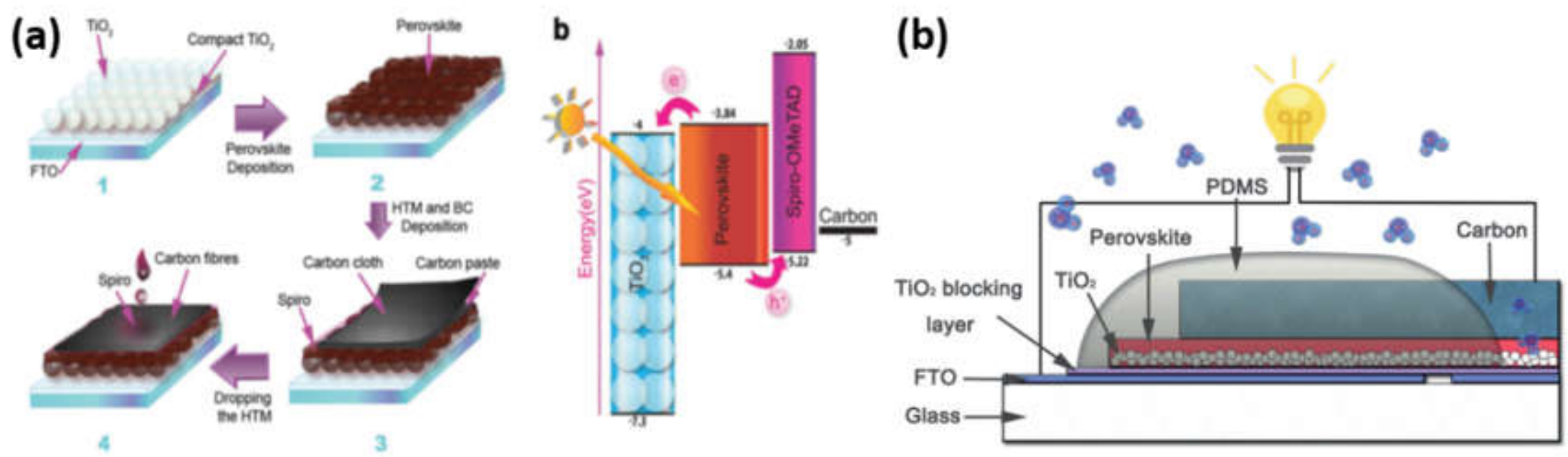

**Fig. 10.** (a) a) A schematic image of the carbon cloth/Spiro-OMeTAD perovskite solar cell fabrication. b) Energy band alignment of a carbon cloth-based device. Reprinted from ref.[42] (b) Schematic representation of the packaged device. Reprinted from ref. [33]

## 2.2. Carbon nanotubes (CNTs)

Carbon nanotubes, also called buckytube, have a particular 1D hollow tubular structure, was uncovered by S.Iijima in 1991.[86] The tubular structure of CNTs can be considered as the result of rolling up monolayer graphite sheet, so CNTs inherit the admirable properties of graphite such as anti-corrosion, impact resistance, thermal and electrical conductivity.[87] Over the past twenty years, CNTs has attracted researchers majoring in the field of electronic and optoelectronic applications due to its exceptional charge transport property, inherent hydrophobicity, chemical and mechanical stability. CNTs have been incorporated into organic solar cells as electron acceptor successfully. Moreover, the p-type CNTs junctures with n-type silicon were demonstrated to achieve a decent PCE up to 17.0%.[88]

As we know, charge selective contacts on both sides of the perovskite absorbers are crucial elements controlling the whole performance of PVSCs. The inherent p-type nature of CNTs and the decent employment as HTM in organic solar cells made CNTs an alternative HTM for PVSCs. One advantage of using CNTs as HTM is eliminating dopants like lithium salts which were blaming to endanger the long-term stability of PVSCs.[87] Also, the mechanical resilience of CNTs may prove to be particularly beneficial for the development of flexible perovskite solar cells. In this section, some significant efforts that are introducing CNTs into PVSCs acting as hole conductor and extraction counter electrode will be presented.

Jang et al. developed a dripping method to drip multi-walled carbon nanotubes (MWCNTs)/chlorobenzene solution on perovskite precursor solution at the time of its spinning, as shown in Fig. 11. The device completed by screen printing carbon paste as counter electrode exhibited a superior efficiency of 13.57% with almost hysteresis-free performance. Compared to cells fabricated by directly spin-coating CNTs on the as-prepared perovskite layer (referred to herein as "post-CNT"), the CNT dripping method played a significant role in the enhanced performance of the PVSCs. Under this modified solvent engineering, CNTs were efficaciously inserted into the grain boundaries of perovskite to create a CNT-perovskite heterojunction. The CNTs shaped motorways for fast hole extraction towards carbon electrode, boosting the $J_{SC}$ value of the devices. Also, contact at the perovskite/carbon interface was enhanced by the CNTs, as is proved by the lessened series resistance ($R_S$) and charge transfer resistance ($R_h$) as shown in Table 3.[89] Additionally, the moisture stability was also enhanced by CNT dripping. For the stability test, the as-prepared devices without encapsulation were exposed at 30 °C under 50% humidity for the first two h and 80% humidity for the next two hours, and the referenced PVSCs employing spiro-OMeTAD as HTM and Au as electrode were tested under the same conditions. The results revealed that the device with CNT modified carbon paste as electrode showed high moisture stability even at 80% RH, retaining 90% initial PCE after the test. By contrast, the reference device lost nearly 80% of original PCE. The author pointed out that the hydrophobic property of the carbon prevented water from permeating into the devices, which is the typical reason for the perovskite degradation. Moreover, in the reference PVSCs, the spiro-OMeTAD as HTM needs to be doped with Li-bis(trifluoromethanesulfonyl)-imide (Li-TFSI) due to poor conductivity of the pristine film. Nevertheless, Li-TFSI is hygroscopic, which is vulnerable to moisture and can accelerate degradation of the perovskite.

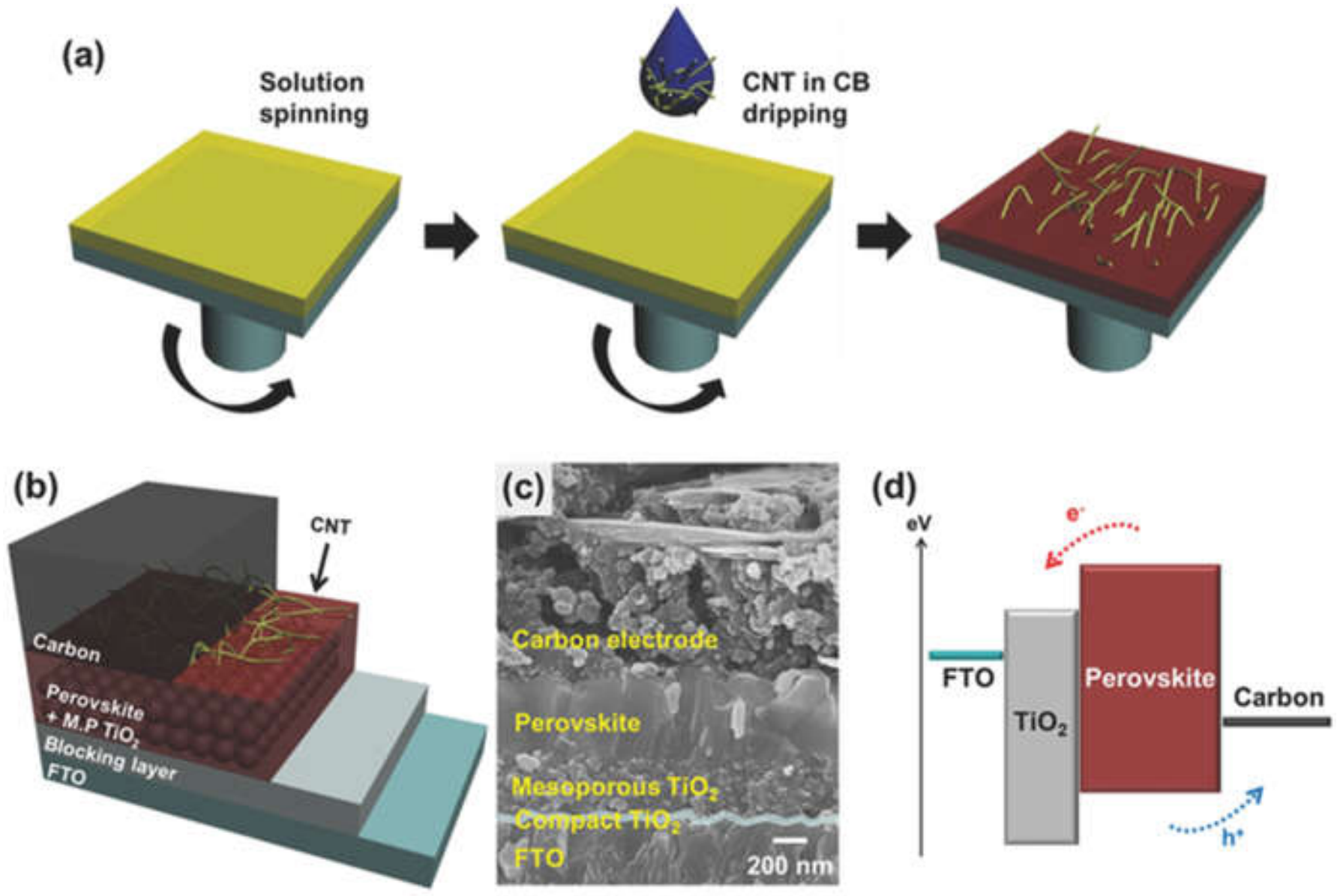

**Fig. 11.** Schematic illustrations of (a) the perovskite layer deposition procedure and (b) the prepared carbon electrode-based perovskite solar cells. (c) A cross-sectional scanning electron microscopy (SEM) image of the prepared device. (d) An energy diagram of the materials used in this work. Reprinted from ref. [89]

Table 3. Summary of resistance values of devices with a perovskite layer fabricated under various conditions. Adapted with permission from ref. [89]

| **Sample** | **$R_S$ [Ω]** | **$R_h$ [Ω]** |
|---|---|---|
| CNT-dripping | 19.04 | 212.46 |
| Post-CNT | 23.01 | 263.97 |

Different from Jang's work that dripping CNTs into perovskite layer, Mingkui Wang et al. designed single-walled carbon nanotubes (SWCNTs)/carbon composite counters electrodes by printing carbon paste on the top of $Al_2O_3$ layer and formed a triple layer structure of $TiO_2$/$Al_2O_3$/C, as shown in Fig. 12(a),(b). This carbon paste was prepared by dissolving SWCNTs, $ZrO_2$, carbon black and graphite into ethyl alcohol. Efficiency up to 14.7% was achieved for device containing 0.05 wt.% SWCNTs in CE with a $V_{OC}$ of about 1.01 V, a $J_{SC}$ of 21.26 mA/cm$^2$, and an FF of 0.69. Under the same conditions, a device without SWCNTs in the CE displayed a PCE of 9.9% with a $V_{OC}$ of 0.889 V, a $J_{SC}$ of 20.26 mA cm$^{-2}$, and an FF of 0.55. Adding SWCNTs in the carbon counter electrode modifies their working function and prolongs the charge recombination lifetime as well. Both are advantageous to the device performance, especially the output photovoltage and FF.[32]

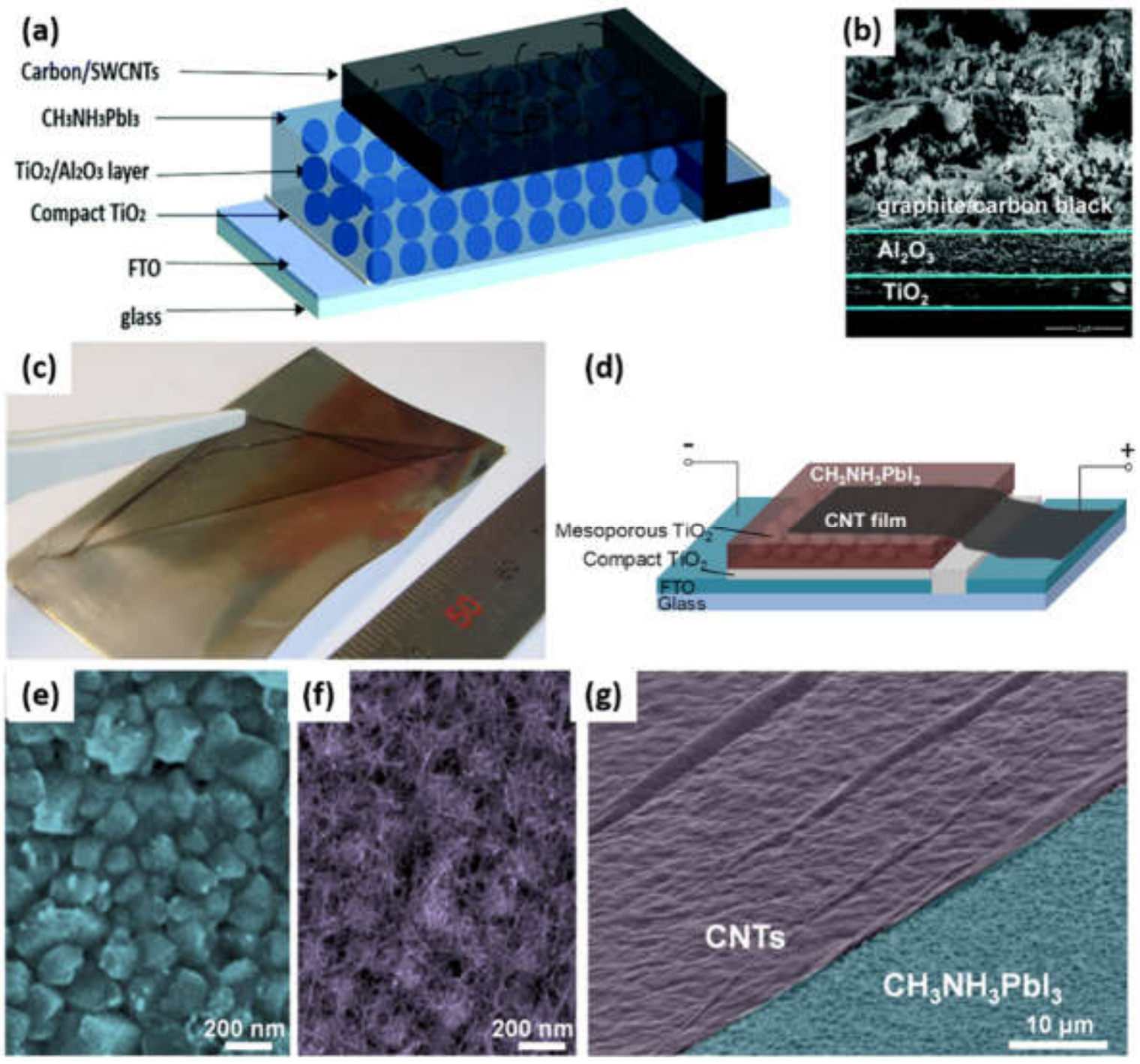

**Fig. 12.** (a) Schematic architecture of the investigated device consisting of FTO as a transparent substrate, a compact $TiO_2$ layer, a mesoporous $TiO_2$ layer, an $Al_2O_3$ layer coated with SWCNT added graphite/carbon black CE, (b) cross-section SEM image of a mesoscopic perovskite solar cell. Reprinted from ref. [32] (c) Photo of freestanding CNT film lifted by tweezers to transfer onto other substrates. (d) Schematic of $CH_3NH_3PbI_3$ perovskite solar cell with CNT film electrode. (e) Top view SEM images of $CH_3NH_3PbI_3$ perovskite substrate before and (f) after CNT transfer. (g) Tilted SEM image of $CH_3NH_3PbI_3$ perovskite substrate (blue) partly covered by CNT film (purple). Reprinted from ref.[66]

One of the initial works that employing CNTs as hole conductor in a mesoscopic PVSCs was made by Mhaisalkar et al. in 2014.[66] In this work, PVSCs was fabricated by laminating freestanding films of CNTs network atop the perovskite absorbers, as shown in Fig. 12(c-g). The network films were synthesized by the floating-catalyst-chemical-vapor-deposition method and then transferred to the as-prepared device easily. It is interesting to note that the CNT film showed a dual-function as both hole-transporter and electrode, thus eliminating the need for a metal electrode. The $CH_3NH_3PbI_3$/CNTs perovskite solar cell without conventional HTM and metal electrode achieved a power conversion efficiency of 6.87% under AM 1.5, 100 mW/cm$^2$ conditions. By adding spiro-OMeTAD to form a composite electrode, the efficiency of this device can be further improved to 9.90% due to the enhanced hole-extraction and reduced charge recombination. With the merits of low cost, a facile fabrication process without a vacuum environment, chemical stability, and electrical compatibility to perovskite, CNTs is potential to replace expensive metal electrodes in PVSCs. Besides, the flexible and transparent CNT film electrodes show great potential in flexible or tandem PVSCs.

Inspired by the work of Mhaisalkar et al., Wong's group used Ti foil as substrate and $TiO_2$ nanotube arrays as ETM to fabricate flexible PVSCs. In Ti foil-based PVSCs, the non-transparent Ti foil impedes the light incidence from the anode, so the device will not operate if using a metallic counter electrode. Therefore, a CNTs network the same as the work above of Mhaisalkar et al. is introduced as the counter electrode. With 25μm Ti foil and $TiCl_4$ treatment to $TiO_2$ nanotube arrays, a decent efficiency up to 8.31% have been achieved. These cells show little deterioration to the device performance after 100 mechanical bending cycles, indicating their excellent flexibility.[67]

Among the family of PVSCs, the clan of carbon-based HTM-free is the most promising for commercialization because of its low-cost and superior long-term stability. However, the FF is lower than that of conventional metal-based PVSCs because of the poor contact in the interface of perovskite/carbon. To address the problem of low FF value which may hinder the way of commercialization, Shihe Yang's group used MWCNT as a hole-selective counter electrode to replace conventional small organic molecule HTM and metallic electrode. With the drop-casting method, graphite, carbon black and MWCNT electrodes were fabricated on the same as-prepared substrates. The performance of three as-fabricated carbon-based PVSCs was demonstrated by the different parameters in Table 4. The scanning electron microscopy (SEM) images of devices demonstrate the existence of the gaps in the interface of graphite/perovskite and the cracks on the carbon black thin film. Nevertheless, the MWCNT not only form a consecutive and crack-free thin film but attach onto the perovskite layer tightly with chemical embedment technique, as shown in Fig. 13(a-f). The morphologies significantly control the contact quality of carbon materials and perovskite layer and impact the device performance. Moreover, the negligible hysteresis effect of the MWCNT-based device compared to the other two was attributed to the excellent interface quality of carbon /perovskite and the matching fast hole-extraction. After optimizing the thickness of $PbI_2$ precursor layer and MWCNT layer, an hysteresis-free efficiency of 12.67% with a very extraordinarily attractive FF of 0.80 were achieved.[72]

Table 4. Solar cell parameters of the three as-fabricated carbon-based PVSCs. Reprinted from ref. [72]

| Solar cells | $V_{oc}$ (V) | $J_{sc}$ (mA/cm$^2$) | FF | PCE (%) |
|---|---|---|---|---|
| Carbon black | 0.90 | 15.98 | 0.65 | 9.35 |
| MNCNT | 0.88 | 15.60 | 0.75 | 10.30 |
| Graphite | 0.93 | 10.30 | 0.64 | 6.13 |

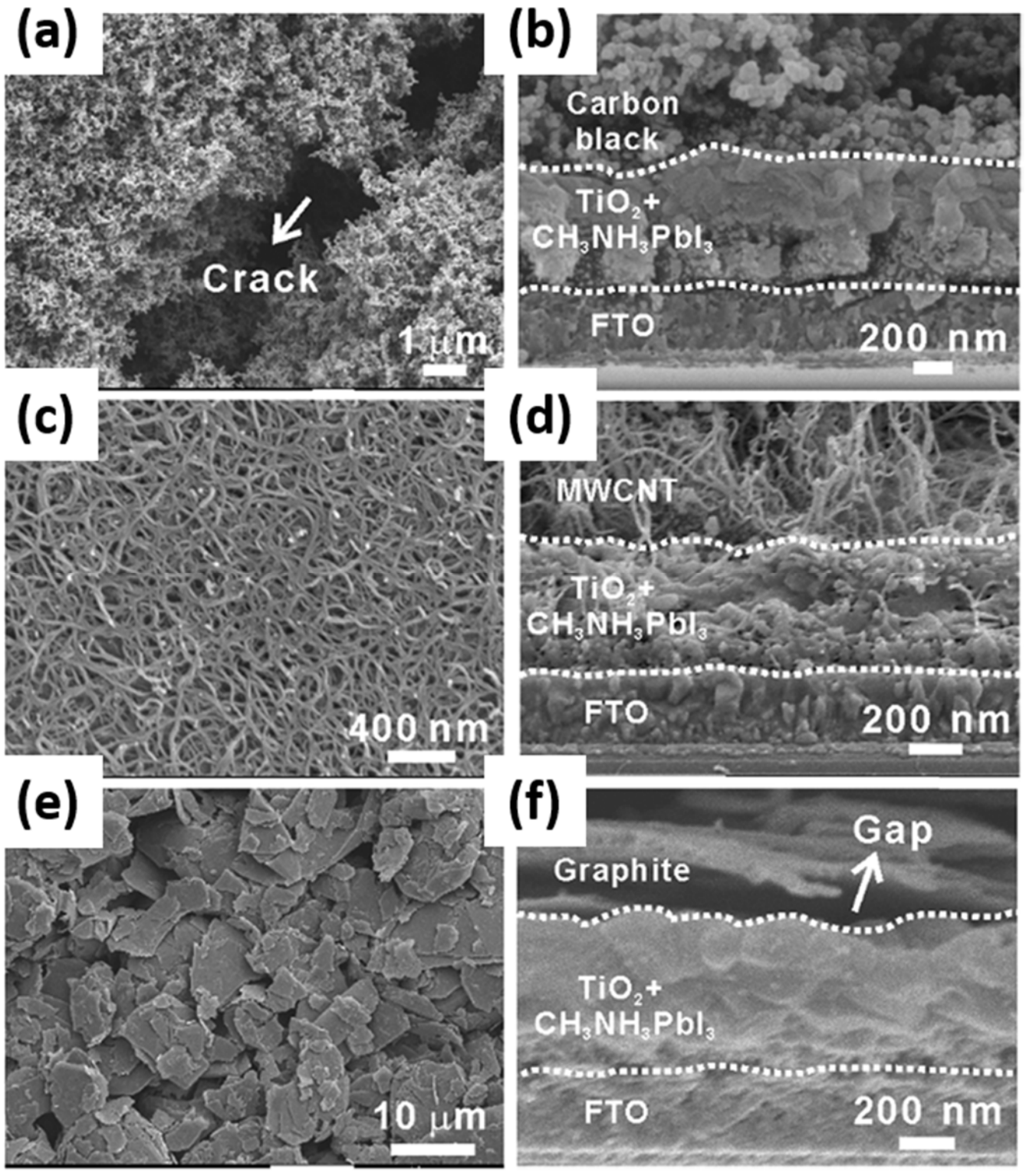

**Fig. 13.** (a, c, and e) Top view and (b, d, and f) cross-sectional SEM images of the carbon black, MWCNT and graphite-based PVSCs. The dashed lines in c, e and g represent the interface of $FTO/TiO_2$ and $CH_3NH_3PbI_3/C$, respectively. Reprinted from ref. [72]

After this, Yang et al. boosted their work by doping boron (B) in the MNCNTs, which increases the work function, carrier concentration, and conductivity of the MWCNTs and enhancing its ability of hole-extraction and transport. Also, an insulating aluminum oxide ($Al_2O_3$) thin layer was coated on the mesoporous $TiO_2$ film as a physical barrier to reduce the charge losses sharply. Finally, the efficiency has been promoted to 15.23%.[71] In another case, Hong Lin et al. integrated a highly conductive and porous cross-stacked super-aligned carbon nanotube (CSCNT) sheet as a metal-free back contact at low temperature in HTM-free mesoscopic PVSCs, as shown in Fig. 14. The device manufactured with manually transferred CSCNT sheets containing 50-layer CNTs (CSCNT-50) presents a PCE of 8.65% due to its three-dimensional hole-extraction and collection. By doping the CSCNT-50 with iodine (I-CSCNT), the PCE was boosted to 10.54% due to the considerable improvement in the FF and $J_{sc}$ because of the increased grain size and improved crystallinity of perovskite as well as the enhanced conductivity of the I-CSCNT electrode.[69]

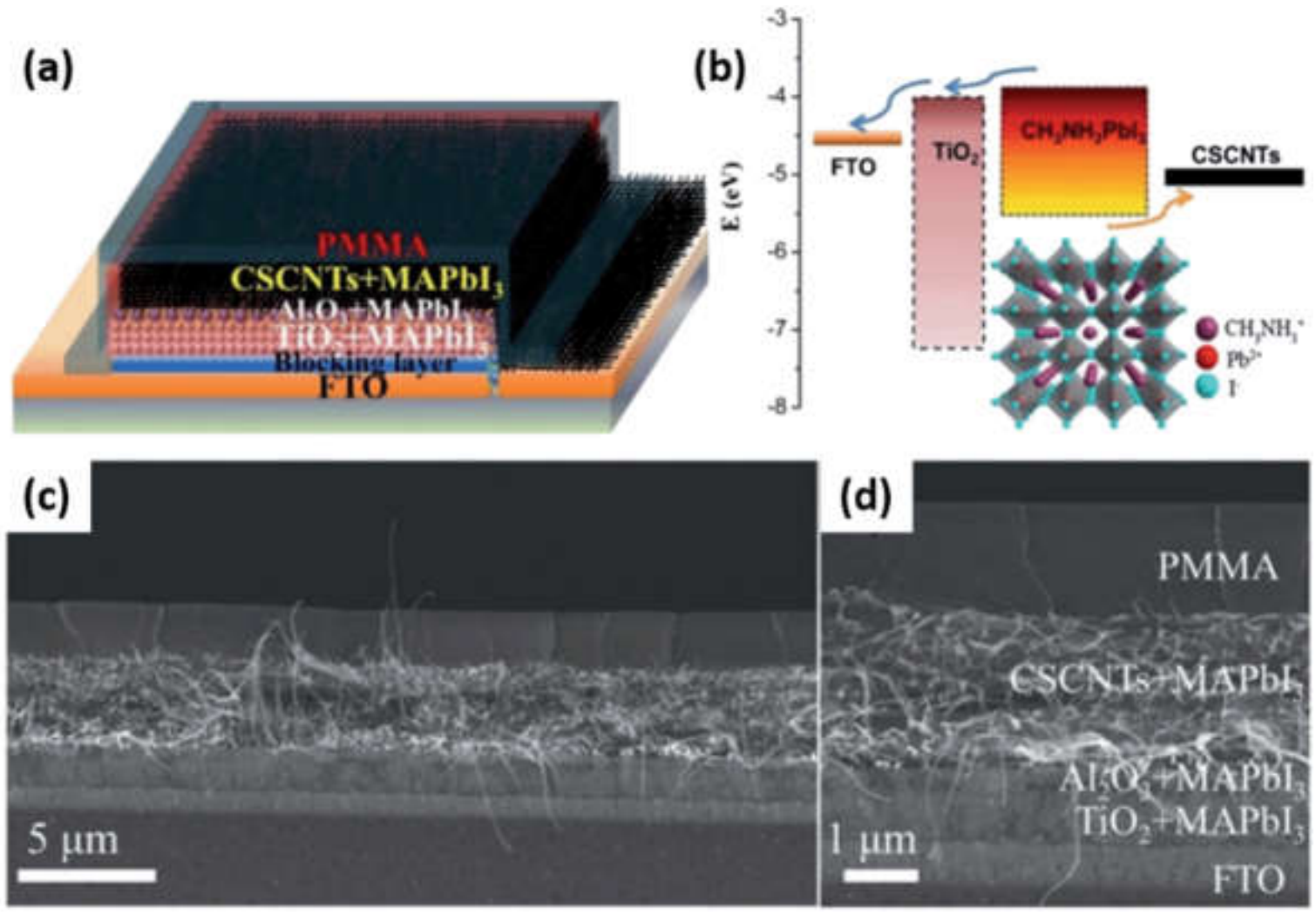

**Fig. 14.** (a) Device architecture of the PVSCs tested in this study; (b) the corresponding energy levels of $TiO_2$, $CH_3NH_3PbI_3$, and CSCNTs; (c and d) the cross-sectional SEM images of the typical CSCNT-based PVSCs with different magnifications. Reprinted from ref. [69]

The overall poor long-term thermal stability is a serious problem that hinders the commercialization of PVSCs. Under elevated temperature condition, considerable amounts of the Au from the electrode diffuse across the HTM into the perovskite layer and cause a dramatic loss of the device performance.[16] Aitola et al. designed a low-cost CNT-based high temperature-stable PVSC with a configuration of FTO/compact $TiO_2$/mesoporous $TiO_2$/perovskite/SWCNT-Spiro-OMeTAD, as shown in Fig. 15. The SWCNT solar cells were prepared by cutting the SWCNT film on filter paper to an appropriate size piece and transferred atop perovskite with some pressure. The SWCNT film was densified with a small amount of chlorobenzene. After that, another SWCNT film was transferred to the first film. The procedure was completed by drop casting Spiro-OMeTAD solution on the SWCNT film. Au-based device with a PCE of 18.4% and SWCNT-based device with a PCE of 15.0% were put under the stability testing. During the stability test with a temperature of 60 °C in the $N_2$ atmosphere and one sun equivalent white LED illumination, the devices showed only a modest linear efficiency loss, which led to an estimated lifetime of 580 h. For comparison, the standard PCSs with Au electrode exhibited a significant and quick efficiency loss mainly due to the ion migration of gold in the structure.[90]

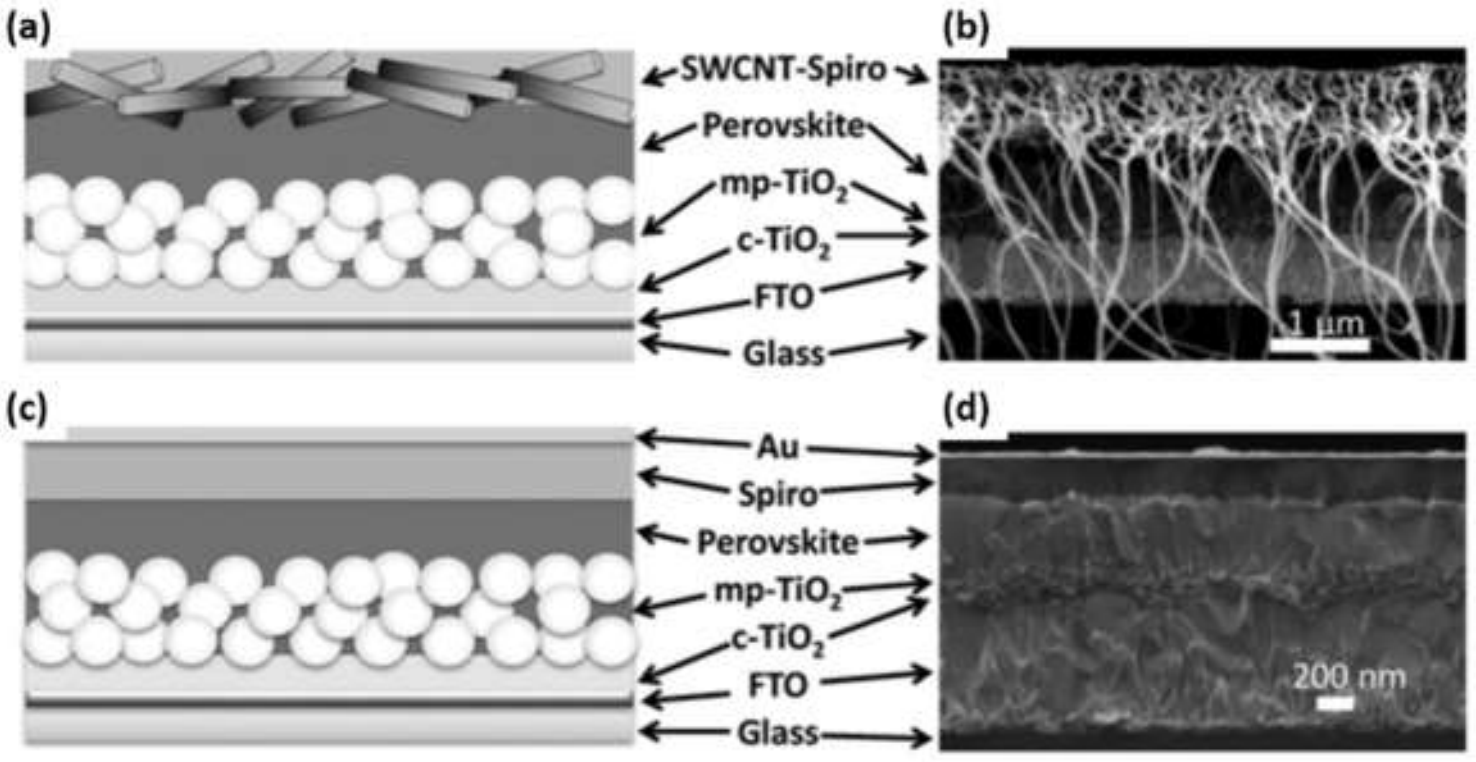

**Fig. 15.** (a) Schematics and (b) cross-sectional SEM image of the SWCNT-contacted device. The carbon nanotube “whiskers” in front of the SEM image are due to sample preparation. (c) Schematics and (d) cross-sectional SEM image of the standard Au contacted device with spin-coated Spiro-OMeTAD HTM. Reprinted from ref. [90]

## 2.3. Graphene

In 2004, Novoselov et al. successfully prepared a monolayer graphene for the first time by mechanical exfoliation.[91] Since then, graphene has attracted numerous researchers in the field of chemical and material science due to its topping mobility, absorbance, conductivity, mechanical flexibility, transparency and specific surface area. Now graphene has played more and more important roles in solar cells, energy storage, optoelectronics, electrics, and sensing.[92–95] The first work of graphene-based PVSCs was reported by Snaith et al. who utilized nanocomposites of graphene and $TiO_2$ nanoparticles as the ETM in PVSCs.[96] One significant application for graphene in PVSCs acting as bottom electrodes may replace conventional ITO or FTO transparent electrodes due to its high conductivity, transparency, charge mobility, chemical and mechanical robustness.[97–99] Also, several studies that are introducing graphene and its derivatives as PVSCs' HTM have been carried out.[65,100–106] Researchers found that perovskite grown on graphene oxide presents a higher orientation order than that on PEDOT: PSS, as affirmed by XRD measurement.[100] Another group found that the use of reduced graphene oxide HTMs with the inherent passivation-ability markedly prolonged the device lifetime compared to PEDOT: PSS based device.[101] Interestingly, Nouri et al. designed graphene oxide as HTM and Li-modified graphene oxide as ETM in PVSCs.[65]

In a word, graphene, a magical material, has given rise to many studies in PVSCs. In this paper, we focus on its role as a counter electrode. PVSCs have been exploited for their application in making transparent or semitransparent devices that can absorb light from both sides for the past few years. Those devices can be applied to the building-integrated photovoltaic system, wearable electronics, and tandem cells. The selections of transparent top electrodes are crucial to the performance of the cells. Different electrodes, such as Au[107–109], Ni[110], CNTs[66], have been studied for their performance in these semitransparent PVSCs. A desired transparent electrode should possess outstanding properties, for instance, high transparency, high conductivity, chemical stability, low cost, and charge collecting ability. Due to its excellent optical and electrical properties, graphene stands out to be an ideal candidate as a transparent electrode in PVSCs. Feng Yan et al. reported a semitransparent PVSCs by laminating stacked multilayer graphene prepared by the chemical vapor deposition (CVD) method as top transparent electrodes, as showed in Fig. 16(a). The graphene electrodes were fabricated and transferred on flexible plastic substrates and then laminated on perovskite active layers at low temperature (≈ 60 ºC) under pressure, which is a facile technique compatible with printing and roll-to-roll processing. The sheet resistance of multilayer graphene will decrease when increases number of graphene layers and contacts with the PEDOT: PSS layer. After optimizing the electrode's conductivity and its contact with HTM, the double-layer graphene electrodes based devices show the average PCEs up to 12.02% and 11.65% when the devices are illuminated from the bottom (FTO side) and top (graphene side) electrodes, respectively.[111]

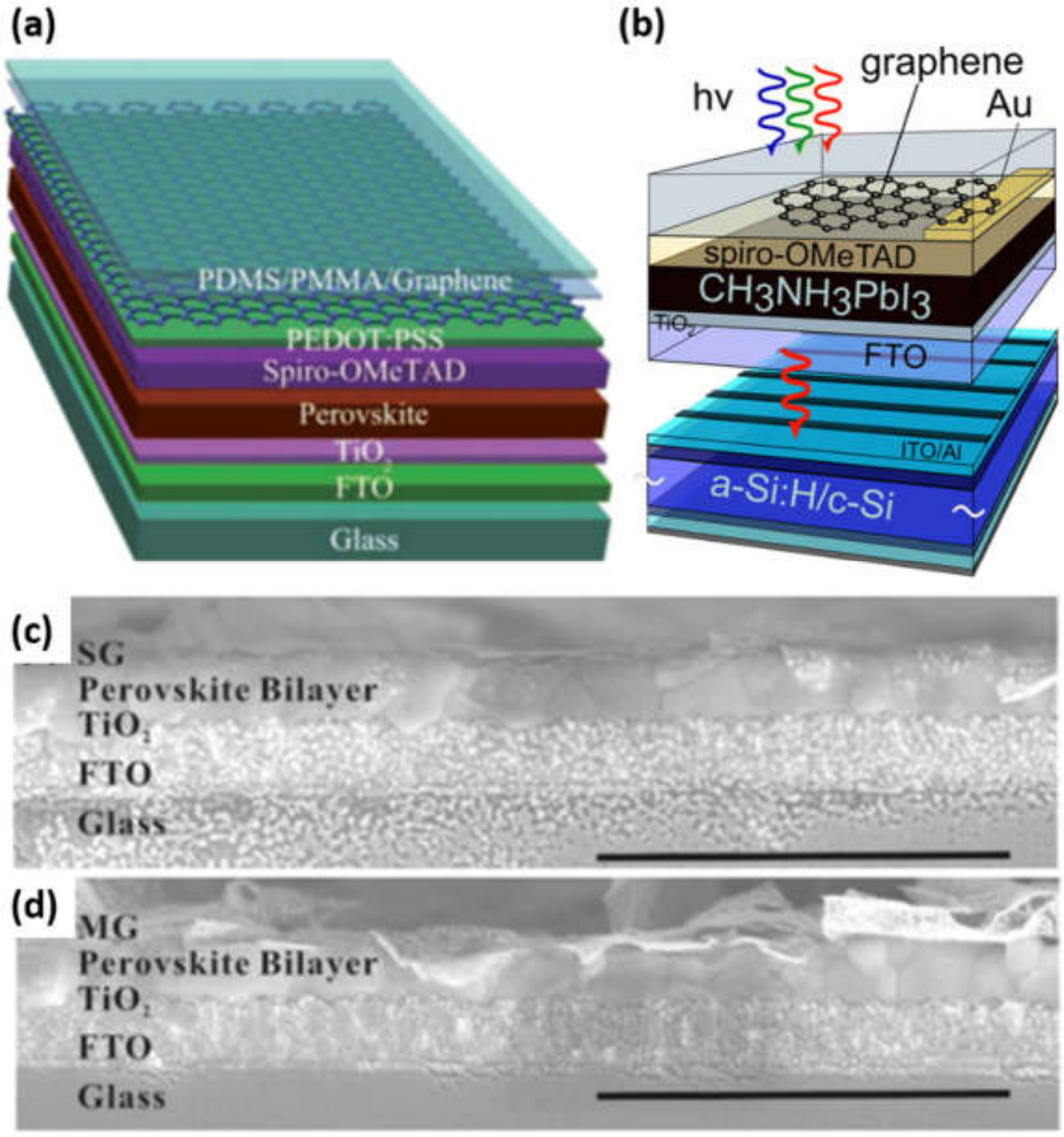

**Fig. 16.** (a) Schematic diagram of a semitransparent perovskite solar cell. Reprinted from ref. [111] (b) Simplified sketch of a four-terminal tandem solar cell consisting of a graphene-based perovskite top solar cell and an amorphous/crystalline silicon bottom solar cell. Reprinted from ref. [112] Solar cell performance using SG and MG for hole extraction. Cross-section view of SG (c) and MG (d) based PVSC (scale bar: 1 μm). Reprinted from ref. [76]

Yang et al. employed flexible reduced graphene oxide to synthesize Single-layered graphene (SG) and multilayered graphene (MG) as hole-extraction electrodes, as shown in Fig. 16(c),(d). Ultraviolet photoelectron spectroscopy(UPS) was used to check the work functions (Fermi level) of the SG and MG materials. Fermi levels of MG and SG were determined at −5.0 eV (similar to graphite) and −4.8 eV, respectively, MG collects the merits of graphite (work function) and monolayer graphene (flexible and spreadable), permitting facile fabrication and energy level engineering simultaneously. Intensity modulated photocurrent/photovoltage spectroscopy (IMPS/IMVS), steady-state photoluminescence (PL) and femtosecond time-resolved photoluminescence (TRPL) revealed that MG-based PVSCs has less recombination and better hole extraction capability than SG-based PVSCs. Indeed, a hole-extraction rate up to 5.1 $ns^{-1}$ measured for MG is comparable to spiro-OMeTAD HTM. Moreover, a Schottky junction was formed in the interface of perovskite/MG, which would accelerate hole-extraction and electron rejection from perovskite. Finally, a decent PCE of 11.5% with a relatively high FF of 0.73 was achieved by using the MG top electrode.[76]

Hu et al. reported a three-dimensional honeycomb-like structured graphene (3DHG) synthesized by the reaction of potassium with $CO_2$.[75] HTM-free PVSCs introduced this 3DHG as counter electrode achieve a PCE of 10.06%. To complete the HTM-free device, 3DHG was coated atop a $CH_3NH_3PbI_3$ layer by doctor blading. Aiming to improve the interface, the 3DHG was dispersed into the solution of perovskite precursors, and the perovskite layer was in situ synthesized. Consequently, a transition layer, in which part of $CH_3NH_3PbI_3$ perovskite molecules was highly scattered by ultrasonic on 3DHG, was formed between the perovskite layer and the 3DHG counter electrode. Common hysteresis effect was not observed for this device, as shown in Table 5.

Table 5. Photovoltaic performance of PVSCs with 3DHG-48 as the CE, Reprinted from ref.[75]

| CEs | $J_{SC}$ (mA cm$^{-2}$) | $V_{OC}$(V) | FF | Efficiency (%) |
|---|---|---|---|---|
| Reverse | 18.11 | 0.89 | 0.63 | 10.06 |
| Forward | 18.86 | 0.86 | 0.61 | 9.84 |

PVSCs with transparent contacts also used to compensate for thermalization losses of silicon solar cells in tandem devices which surpass the Schockley-Queisser limit for single-junction solar cells. PVSCs with large-area chemical vapor deposition (CVD)-graphene was fabricated by Nickel et al. for Tandem Solar Cells, as shown in Fig. 16(b).[112] The performance of the device with a configuration of glass/FTO/$TiO_2$/$CH_3NH_3PbI_3$/spiro-OMeTAD/graphene is comparable to those of PVSCs with regular Au contact. The PCE of the tandem device was 13.2%, thus increased by 30% as compared to the respective perovskite single-junction cell.

## 3. Carbon Materials for Large Area Solar Module

Most of the devices above have small active area of around 0.1 cm$^2$. However, as a market-proof photovoltaic technology solution, one needs to demonstrate solar modules with larger area. To make solar panels with conventional structure employing 'layer-by-layer' process, it is challenging to obtain large-area, flat, uniform and fully covered perovskite film. To solve the problem, we deposited perovskites by slot-die coating technique assisted with our previously reported gas-pumping method [113,114]. (Fig. 17(a),(b)). In the end, carbon paste was screen-printed on top of the perovskite film as counter electrode. The as-prepared 5*5 cm$^2$ module showed a PCE of 10.6% and good stability over 140 days of outdoor testing. [115]

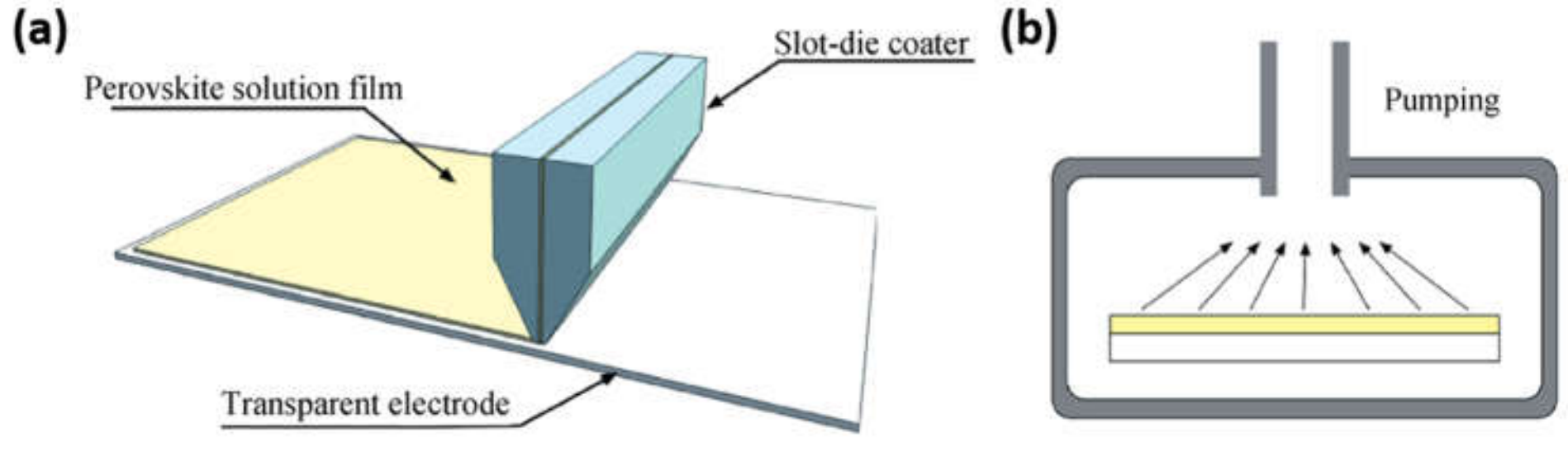

**Fig. 17.** (a) Schematic diagram of slot-die coating and (b) pumping of perovskite liquid film. Reprinted from ref. [115]

In 2016, Mhaisalkar et al. reported monolithic perovskite modules with active areas of 70 $cm^2$. The module was fabricated by screen printing along with infiltration of the perovskite $(5\text{-AVA})_x(MA)_{1-x}PbI_3$ solution, showing a PCE of 10.74% and good ambient stability of over 2000 h with less than 5% reduction in efficiency. They pointed out that the quality of mesoscopic carbon had a significant impact on the module performance, where high BET (Brunauer–Emmett–Teller) surface area and low electrical resistance are propitious for achieving an efficient module.[116]

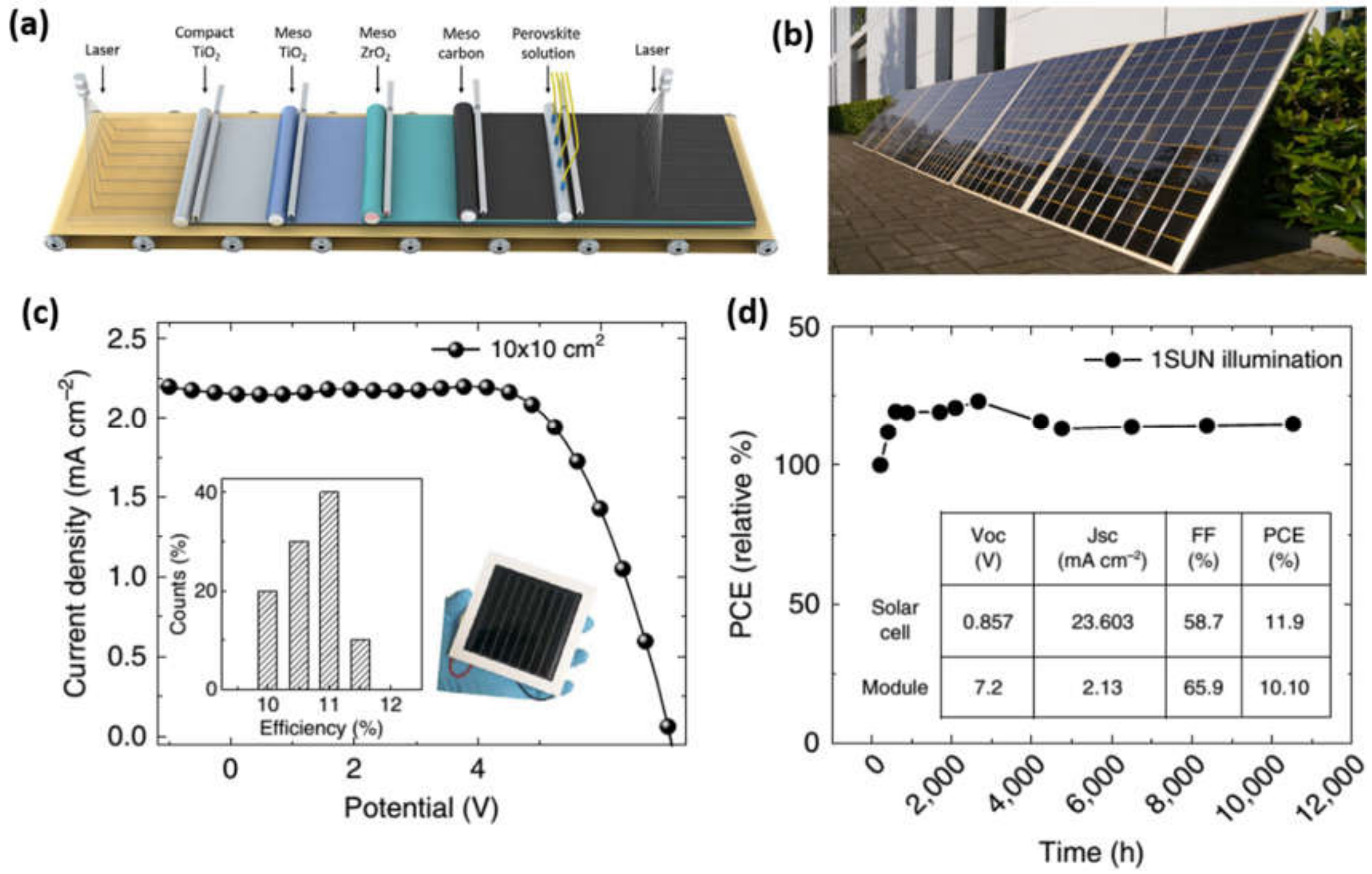

**Fig. 18.** (a) Schematic illustration of the proposed production line of perovskite solar modules. (b) Image of 7 $m^2$ printable perovskite solar panels. Reprinted from ref. [117] 2D/3D Carbon based Solar cell characteristics and stability. (c) J-V curve using the 2D/3D perovskite with 3%AVAI in an HTM-free 10*10 $cm^2$ module (device statistics and picture in the inset). (d) Typical module stability test under 1 sun AM 1.5 G conditions at stabilized temperature of 55 °C and short circuit conditions. Stability measurements are done according to the standard aging conditions. Reprinted from ref. [118]

Later, employing the same structure, Han et al. optimized the thickness of the mesoporous $TiO_2$ and $ZrO_2$ layer and fabricated large-area module (active area of 49 $cm^2$ out of total 100 $cm^2$) with a PCE of 10.4%. The module was stable under light soaking test with AM1.5 illumination for 1000 h and tested for a month under outdoor conditions. The shelf life under ambient conditions in the dark was over 1 year. Moreover, to assess the reproducibility of their screen-printing technique in manufacturing high-performance perovskite solar modules, a 7 $m^2$ fully printed perovskite solar panel was also demonstrated.[117] (Fig. 18(a),(b))

Recently, Nazeeruddin et al. successfully fabricated 10*10 $cm^2$ (active area of 47.6 $cm^2$) monolithic structure modules with a fully printable industrial-scale process, showing the efficiency of 11.2%. More importantly, the sealed module kept stable for one-year (> 10,000 h) under 1 sun AM 1.5 G illumination conditions at a stabilized temperature of 55 °C. (Fig. 18(c),(d)). They ascribed their ultra-stable solar cells to using mixed-cation perovskite of $(5AVA)_2PbI_4$ and $CH_3NH_3PbI_3$, which could form 2D/3D interface in the devices. Moreover, the 2D perovskite $(5AVA)_2PbI_4$ acts as a protective window against moisture, preserving the 3D perovskite.[118] These reports on large-area modules demonstrated the potential of carbon-based PVSCs to make up-scale production and paved the way to commercialization of perovskite solar cells.

## 4. Summary and Outlook

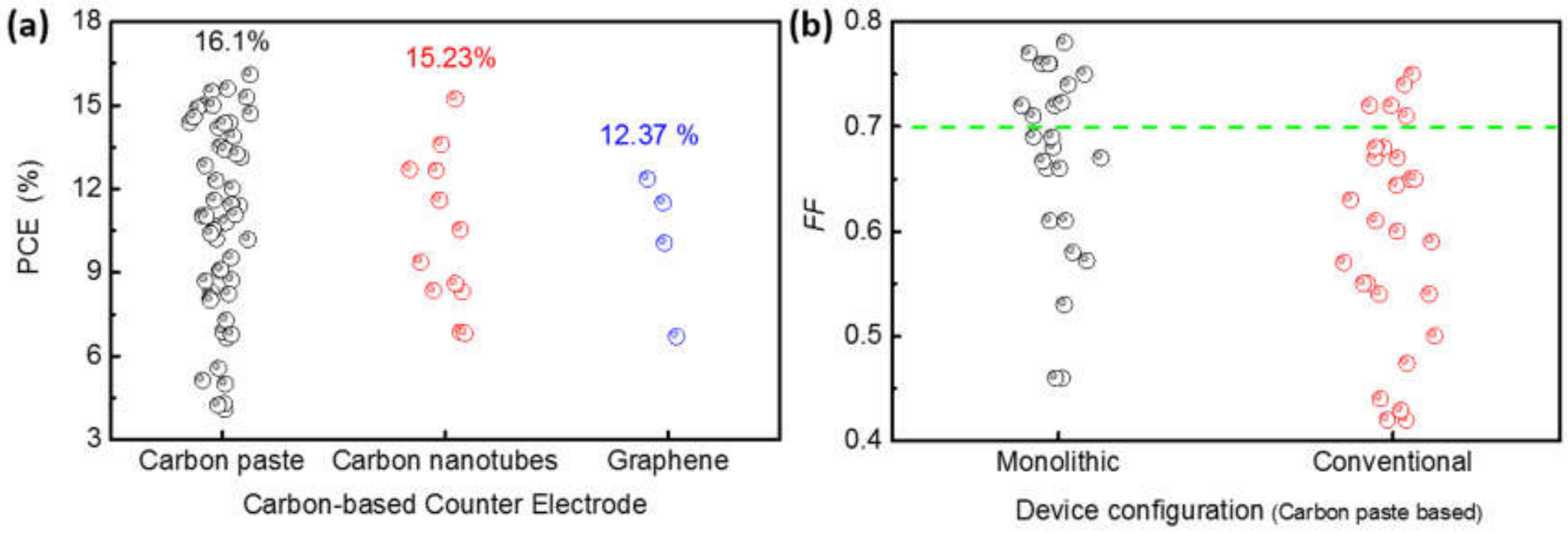

**Fig. 19**. Comparison of the performance of the device between (a) different types of carbon-based counter electrode. (b) different types of device structures. (data were analyzed based on table 1)

CE is a crucial component of PVSCs, in this review, we introduce some critical works in the field of PVSCs that focus on designing or modifying cost-effective carbon-based CEs. After summing up this works, we found that there are mainly three key factors, included electrical conductivity, the closeness of the perovskite (or HTM)/CE interface and matching work function that determines the performance of CEs and therefore the overall performance of the devices. Carbon materials attracted quite a few researchers because of their earth-abundant, low-cost and decently conductive properties. Carbon materials can be deposited on the top of the as-prepared substrate by different methods such as screen printing, doctor-blading, inkjet-printing, rolling transfer, hot pressing, spin coating, and press transferring, and drop-casting. Although this kind of CEs often meets poor contact with perovskite or HTM, different approaches such as doping, morphology tuning, and device engineering have been used to alleviate this problem and improve the performance of carbon-PVSCs. Among the carbon materials, CNTs and graphene stand out to be potential candidate CEs for semi-transparent or transparent PVSCs, which can be employed in building-integrated photovoltaics and tandem solar cells. Moreover, device lifetime up to 3,000 hours has been reported based on carbon cell. A comparison of the device performance between different types of carbon-based CEs is visualized in Fig. 19(a). Firstly, it is clear that more efforts were put on the use of standard carbon paste due to the easy availability and low cost. The highest PCE was also achieved from the device with carbon paste based CEs. Relatively lower PCEs and fewer reports on the CNT and Graphene based devices, which could be due to the emphasis on the transparency of these types of solar cells. In Fig. 19(b), we attempted to plot the FF values from two main kinds of devices based carbon pastes: Monolithic and Conventional. Generally, the former structure has higher FF than the latter, possibility due to better contact and lower resistance of the monolithic structures. By the way, the progress of large-area modules based on carbon CEs are presented, and the potential of scaling up PVSCs with carbon CEs was confirmed.

It is evident that there is still a huge gap in PCEs between the carbon-based devices and noble metal-based devices. There is also an ample space for improving the device performance if we can further increase the conductivity of the carbon electrode and improve the contact between the electrode and charge-selective contacts. On the other hand, traditional PVSCs employing noble metals as CEs usually need organic HTMs such as spiro-OMeTAD, PTAA, and PEDOT: PSS, which deteriorate device stability under high humidity or high-temperature conditions. However, the carbon-based devices such as monolithic devices or devices using CNTs as HTM can avoid the use of these fragile organic compounds, and therefore the moist and thermal stability can be enhanced. More importantly, a relatively large-area module with a lifetime up to 10,000 hours has been reported based on HTM-free carbon cells. Thus, given the low cost and long-term stability of the carbon materials, they can be an essential solution for the commercialization of perovskite solar cells.

**Acknowledgment**

P.G. acknowledges the financial support from “Hundred Talents Program” of the Haixi Institute Chinese Academy of Sciences (NO. 1017001) and “Thousand Talents Program” of the Government of China.